\documentclass{jfm}
\usepackage{graphicx}
\usepackage{epstopdf, epsfig}

\usepackage[justification=justified,format=plain, width=\textwidth]{caption}
\usepackage{hyperref}
\hypersetup{
    colorlinks = true,
    urlcolor   = blue,
    citecolor  = blue,
}
\usepackage{xcolor}

\shorttitle{Small-scale intermittency of flames}
\shortauthor{Roy et al.}

\title{Small-scale intermittency of premixed turbulent flames}

\author{Amitesh Roy\aff{1}
  \corresp{\email{amiteshroy94@yahoo.in}}\aunote{Presently at the Institute for Aerospace Studies, University of Toronto, Ontario, Canada},
  Jason R. Picardo\aff{2}\corresp{{Also, Associate, International Centre for Theoretical Sciences, TIFR, India.}},
  Benjamin Emerson\aff{3},
  Tim C. Lieuwen\aff{3}
 \and R. I. Sujith\aff{1}}

\affiliation{\aff{1}Department of Aerospace Engineering, IIT Madras, Chennai 600 036, India
\aff{2}Department of Chemical Engineering, IIT Bombay, Mumbai 400 076, India
\aff{3}Daniel Guggenheim School of Aerospace Engineering, Georgia Institute of Technology, Atlanta, Georgia, 30332, USA}
\begin{document}

\maketitle

\begin{abstract}
Premixed turbulent flames, encountered in power generation and propulsion engines, are an archetype of a randomly advected, self-propagating surface. While such a flame is known to exhibit large-scale intermittent flapping, the possible intermittency of its small-scale fluctuations has been largely disregarded. Here, we experimentally reveal the inner intermittency of a premixed turbulent V-flame, while clearly distinguishing this small-scale feature from large-scale outer intermittency. From temporal measurements of the fluctuations of the flame, we find a frequency spectrum that has a power-law subrange with an exponent close to $-2$, which is shown to follow from Kolmogorov phenomenology. Crucially, however, the moments of the temporal increment of the flame position are found to scale anomalously, with exponents that saturate at higher-orders. This signature of small-scale inner intermittency is shown to originate from high-curvature, cusp-like structures on the flame surface, which have significance for modeling the heat release rate and other key properties of premixed turbulent flames.
\end{abstract}


\section{Introduction}
\label{sec:intro}
The dynamics of a flame in a turbulent pre-mixture of fuel and oxidant is of central importance to combustion processes and plays a key role in present-day power generation and propulsion engines. The fluctuating motion of the flame surface, which separates burned from unburned gases, is the result of a complex interplay between the propagation speed or burning velocity of the flame (which is determined by its inner chemical structure) and the multi-scale turbulent velocity field of the carrier flow \citep{peters2001turbulent, driscoll2008turbulent, lipatnikov2010effects, steinberg2021structure}. The outcome is a fractal flame surface with spatial fluctuations spanning a wide range of length scales, from the size of the system down to the scale of dissipative processes \citep{gouldin1987application,peters1988laminar,gulder2000flame,chatakonda2013fractal}. In between lies an apparently self-similar range wherein the flame fluctuations have a Fourier spectrum that varies as a power-law, which in turn follows from the inertial-range scaling of the underlying turbulent flow \citep{peters1992spectral, peters2000modification, chaudhuri2011spectral, chaudhuri2012flame}. The statistical properties of the flame surface are widely recognized to determine crucial quantities such as the turbulent flame speed, as well as the rates of reaction, and volumetric heat generation \citep{kerstein1988field, lipatnikov2002turbulent, chaudhuri2012flame}. Despite this, previous studies have almost entirely overlooked an essential property of the power-law subrange of fluctuations, namely its intermittency.

This small-scale, \textit{inner} intermittency is fundamentally different from the well-studied \textit{outer} intermittency of the large-scale flame motion, which is characterized by on-off flapping \citep{bray1985unified, robin2011direct, cheng1987intermittency, poinsot2005theoretical}. Although the need to recognize and study inner intermittency was emphasized by \citet{sreenivasan2004possible}, the literature remains sparse \citep{kerstein1991fractal, gulder2007contribution, roy2020fractal} with no experimental work, leaving a fundamental gap in our understanding of turbulent flame dynamics.

In this paper, we experimentally uncover the inner intermittency of a turbulent, $\mathrm{CH}_4$-air V-flame, using high-frequency temporal measurements of the flame surface. We first clearly distinguish outer intermittency, which is apparent from the probability distribution function (PDF) of flame fluctuations $\xi^\prime$, from inner intermittency, which is only revealed after a scale-by-scale analysis using temporal increments of the flame position $\delta\xi^\prime(\tau)$. As the time interval $\tau$ decreases, the PDFs of $\delta\xi^\prime$ exhibit increasingly non-Gaussian, flared tails indicating that the flame fluctuations contain rapid extreme events. 

Next, we show how the apparently self-similar range, hitherto studied primarily in spatial wavenumber space, manifests in the temporal frequency domain: the spectrum of ($\xi^\prime$) has a power-law subrange, with an exponent that is shown to agree with Kolmogorov phenomenology. However, when we analyze the structure functions (moments of $\delta\xi^\prime$) we find that they scale anomalously, with exponents that saturate at high orders. Thus, we show that the small-scale flame fluctuations violate perfect self-similarity and are, in fact, strongly intermittent. Moreover, the extreme-values of $\delta\xi^\prime$ are found to originate from the advection of high-curvature, cusp-like structures along the flame surface. These findings have important implications for the modelling of premixed turbulent flames and suggest new directions for future work, as discussed in the concluding section of this paper.

Intermittency in \textit{non-reacting} turbulent flows has been well studied, with comparable attention paid to both varieties. Outer intermittency is encountered in the large-scales of non-homogeneous or transitional flows \citep{kovasznay1970large, avila2011onset, barkley2015rise}, while inner intermittency is a characteristic feature of the inertial range of fully-developed, homogeneous, isotropic turbulence \citep{frisch1995turbulence, sreenivasan1997phenomenology,arneodo2008universal}. The intermittency of a passive and conserved scalar field, stirred by a turbulent flow, has also been studied in detail \citep{holzer1994turbulent, tong1994passive, warhaft2000passive}, in part as a first step towards understanding the intermittency of turbulence \citep{shraiman2000scalar, falkovich2001lag, Falkovich2006lessons}: the concentration fluctuations remain intermittent even when the turbulent flow is replaced by a simpler, non-intermittent, Gaussian flow \citep{shraiman2000scalar,Tsinober2009}. Intermittent scalar fields are characterized by sharp internal fronts or ramp-cliff structures, across which the scalar experiences the largest possible fluctuation over the smallest (diffusive) spatial scale \citep{celani2000universality, watanabe2006intermittency}. 

\textcolor{black}{In contrast, in the \textit{combustion} literature, inner intermittency of scalar fields---which are neither passive nor conserved---has remained largely neglected, save for a few studies. Importantly, ramp-cliff structures have been experimentally observed in the scalar fields underlying partially premixed turbulent flames \citep{wang2007experimental, cai2009investigation}. 
} Also, the extreme-value statistics of the dissipation range have been characterized, for example, through log-normal distributions of scalar dissipation, in reacting flow experiments \citep{karpetis2002measurements,saha2014flame} and simulations \citep{hamlington2012intermittency, chaudhuri2017flame}. \textcolor{black}{However, a clear characterization of inertial range inner intermittency in the dynamics of the flame surface is missing, and this is the focus on our work.}

\section{Experimental setup}
\label{sec:expt}

\subsection{Facility}
\begin{figure}
{\centering
\includegraphics[width=\textwidth]{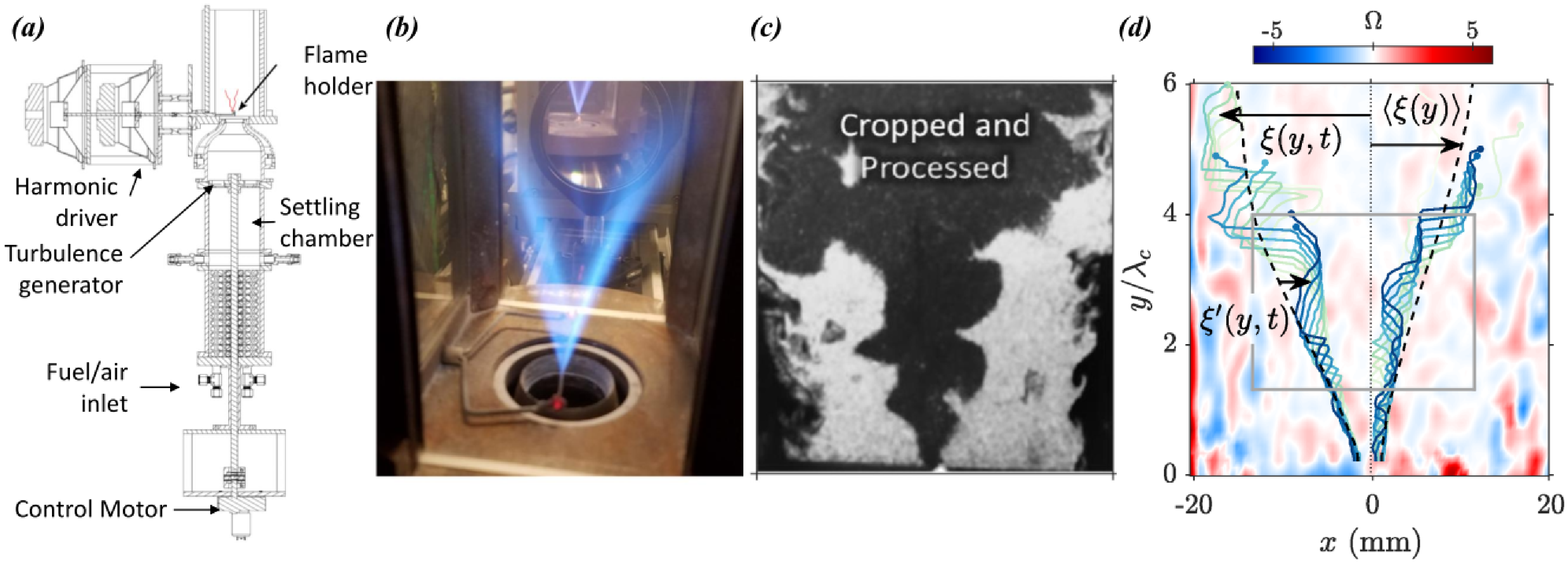}}
\caption{\label{Fig1} Turbulent V-flame facility. (a) Schematic of the combustor setup. (b) Illustrative photograph of the V-flame. (c) Representative Mie scattering snapshot depicting the flame surface. (d) Examples of the extracted instantaneous flame edge $\xi(y,t)$, along with the coordinate system. The dashed line represents the mean flame edge $\langle \xi(y)\rangle$ with respect to which the fluctuations $\xi^\prime(y,t)$ are defined. The region highlighted by the box is used for determining the flow properties. Panels (a-c) adapted from \cite{humphrey2018premixed} with permission from Cambridge University Press.}
\end{figure}

Our experimental facility (figure \ref{Fig1}a) consists of a premixed $\textrm{CH}_4-$air V-flame stabilized on an oscillating flame holder, a typical configuration for the study of premixed flames \citep{petersen1961stability}. The flame holder, which is an electrically heated nichrome wire, is vibrated at a frequency of $f_f=1250$~Hz. The flow of $\textrm{CH}_4$ and air ensues out of a circular nozzle 10 mm below the flame holder, and turbulence is generated using a series of stator-rotor plates. \textcolor{black}{The three-dimensional (3D) flame surface, thus obtained, is shaped like a wedge (figure \ref{Fig1}b). We measure the fluctuations of the flame edge and the underlying velocity field within a two-dimensional plane located at the mid-section of the wedge-shaped flame surface, using $\textrm{TiO}_2$ Mie scattering and high-speed particle image velocimetry (PIV).} This experimental setup has been employed previously for studying the effects of harmonic forcing and turbulence on premixed flames~\citep{humphrey2018premixed, roy2019nonlinear}.

\subsection{Flame and flow characteristics}
\label{sec:flameprop}
We consider two experimental flame configurations, F1 and F2, whose properties are listed in table~\ref{tab1}. For both flames, the turbulence intensity is $u^\prime/\bar{u}_y \approx 0.1$, where $u^\prime$ is the root-mean-square (r.m.s.) of the velocity fluctuations and $\bar{u}_y$ is the mean longitudinal velocity. The Reynolds number $Re_\ell=\ell u^\prime/\nu$ of F1 is about 700 while that of F2 is nearly four times larger; the dissipative Kolmogorov length ($\eta$) and time ($\tau_\eta) $ scales are similar though for both flames (cf.~table~\ref{tab1}). 

The flame speed $s_L$ 
is calculated using Chemkin Premix \citep{kee2011chemkin} with detailed chemistry simulated through the GRIMech 3.0 mechanism \citep{smith1999gri} at 300 K and 1 bar. \textcolor{black}{The associated 
Karlovitz number $Ka =(\nu/Sc\, s_L)^2 \eta^{-2} \sim 0.1$, for both configurations, which implies that the flames lie within the corrugated flamelet regime, close to the boundary with the thin reaction zone regime \citep{peters2001turbulent}.} Thus, the flame front is  continuous, enabling a well-defined description of the flame edge.

\textcolor{black}{Turbulent eddies can distort the flame edge and produce wrinkles or corrugations, but on scales on the order of the Gibson scale $\ell_g  = (s_L/u^\prime)^3\ell \approx 0.3$ mm or greater. This is because smaller eddies have velocities less than the laminar flame speed $s_L$ and so cannot distort the flame edge~\citep{peters2001turbulent}.}
Further information on the properties of the flame and the flow, as well as on how these are calculated, is provided in the Supplemental Materials.

\begin{table}
  \begin{center}
\def~{\hphantom{0}}
{\color{black}
\begin{tabular}{lccc}
\textbf{Property} & \textbf{Symbol} & \textbf{Flame F1} & \textbf{Flame F2}\\[3pt]
      equivalence ratio & $\phi$ & $0.97$ & $0.91$ \\
      mean longitudinal velocity & $\bar{u}_y$ & $5.1$ m/s & $7.33$ m/s\\
      r.m.s. velocity fluctuation & $u^\prime$ & $0.59$ m/s  & $0.97$ m/s\\
      turbulence intensity & $u^\prime/\bar{u}_y$ & $0.12$ & $0.13$ \\
      kinematic viscosity & $\nu$ & $2.29\times10^{-6}$ m$^2$/s & $2.27\times10^{-6}$ m$^2$/s\\
      flow integral length scale & $\ell$ & $2.72$ mm & $6.22$ mm\\
      flow integral time scale & $\tau_\ell=\ell/u^\prime$ & $4.61\times10^{-3}$ s & $6.41\times10^{-3}$ s\\
      Reynolds number & $Re_\ell=u^\prime\ell/\nu$ & $701$ & $2670$\\
     Kolmogorov length scale & $\eta=Re_\ell^{-3/4}\ell$ & $0.020$ mm & $0.017$ mm \\
     Kolmogorov time scale & $\tau_\eta=Re_\ell^{-1/2}\tau_\ell$ & $1.75\times10^{-4}$ s & $1.24\times10^{-4}$ s \\
     Schmidt number & $Sc$ & 0.7 & 0.7 \\
     Corrsin length scale & $\eta_c=Sc^{-3/4}\eta$ & $0.026$ mm & $0.022$ mm\\
     flame speed & $s_L$  & $0.37$ m/s & $0.34$ m/s \\
     Karlovitz number & $Ka=(\nu/Sc\, s_L)^2 \eta^{-2}$ & $0.20$ & $0.32$\\
     Gibson length scale & $\ell_g=(s_L/u^\prime)^3\ell$ & $0.68$ mm & $0.27$ mm\\
\end{tabular}
}
\caption{
Relevant physical properties of the two turbulent premixed flame configurations considered in this study. The laminar flame speed $s_L$ for the two cases was obtained using Chemkin Premix calculations \citep{humphrey2017ensemble}, while the value of $Sc$ for methane-air premixed flames was obtained from \cite{tamadonfar2014flame}. See the Supplemental Materials for further details.\label{tab1}}
\end{center}
\end{table}

{
\color{black}

\subsection{Window of interrogation}
\label{sec:window}
Our analysis of small-scale flame dynamics is based on measurements in a two-dimensional plane, within a sub-region of the entire domain (outlined by the grey rectangle in figure \ref{Fig1}d). The extent of the sub-region in the longitudinal $y$-direction is chosen such that the flame fluctuations are not dominated by effects of flame anchoring and the oscillation of the flame holder, which is ensured for $y>\lambda_c$, where $\lambda_c=\bar{u}_y/f_f$ (notice the disappearance of the narrow-band peak from the power spectra in figure \ref{Fig3}). Further, we disregard fluctuations at large downstream distances ($y>4\lambda_c$) where effects of large-scale flapping become significant [discussed further below in connection with figure \ref{Fig2}(a,b)]. The width of the sub-region in the transverse $x$-direction is restricted by the requirement that the measured velocity fluctuations exhibit nearly isotropic statistics. This is verified by ensuring that the cross-correlation $\langle u_x^\prime u_y^\prime\rangle$ remains small (see \S3 in Supplemental Materials).

Our measurements in the $x-y$ plane give us access to the fluctuations of the flame edge in the direction normal to the mean flame edge (dashed line in figure~\ref{Fig1}d), as well as in the flow-aligned tangential direction. However, we do not have access to the fluctuations in the out-of-plane tangential direction ($z$-direction). We do not expect this omission to affect out key results, though, because within the sub-region of interest where the flow is approximately isotropic the only difference between these two tangential directions is the advection by the mean flow in the flow-aligned direction. We can account for this advection using Taylor's hypothesis ($u^\prime/\bar{u}_y \sim 0.1$) and thereby approximate the $z$-direction fluctuation statistics from data of the flow-aligned tangential fluctuations \citep{shin2013flame}. This procedure is facilitated by the local homogeneity in the $z$-direction near the mid-section of the wedge-shaped flame surface where the measurement plane is located. In fact, many studies have carried out similar two-dimensional measurements for estimating important flame properties such as the fractal dimension of the flame surface \citep{north1990fractal, smallwood1995characterization, gulder2000flame}.

\subsection{Spatial and temporal resolution}
\label{sec:resol}
A laser sheet of thickness $1$ mm is used for Mie-scattering and PIV. The resulting images capture a region spanning $50\times 60$ mm$^2$, and the size of a pixel is $\Delta x=0.078$ mm. At this resolution the flame edge appears as a distinct boundary in the processed Mie-scattering images (figure~\ref{Fig1}c). The temporal frequency at which the images are obtained is $f_s=1.25\times10^4$ Hz, which based on the Nyquist theorem allows us to capture fluctuations of the interface with a maximum frequency of $f_{\textrm{max}}=f_s/2 = 6.25\times10^3$ Hz, corresponding to a time interval of $1/f_{\textrm{max}}=1.6\times10^{-4}$s.

This spatio-temporal resolution is just sufficient to resolve the fluctuations of the flame edge. The thickness of the laser sheet is of the same order as the Gibson scale $\ell_g$ (\S~\ref{sec:flameprop}), which is an estimate of the scale of the smallest turbulence-induced wrinkles on the flame edge \citep{peters2001turbulent}. The pixel width $\Delta x$ is an order smaller than $\ell_g$, and so we are able to capture the spatial undulations of the flame edge.  At the Gibson scale, the inertial-range turbulent eddies have a time scale of $\ell_g/(u^\prime (\ell_g/\ell)^{1/3})\approx 10^{-3}$ s (following Kolmogorov phenomenology). An even smaller convective time-scale is obtained by considering the advection of small spatial undulations along the flame surface past a fixed measurement location; using the mean longitudinal velocity as an upper estimate for the speed of convection, we obtain a time-scale of $\ell_g/\bar{u}_y \approx 10^{-4} \mathrm{s}$ which is approximately the same as the smallest resolved time-scale $1/f_{\textrm{max}}$. The ability of our measurements to capture such events will turn out to be especially important for detecting the inner intermittency of the flame dynamics (cf.~\S~\ref{sec:curvature}).

While we can analyse the flame fluctuations in detail, our resolution is  insufficient to capture the dynamics of the underlying turbulent flow field. Indeed, while the pixel dimension is of the order of the Kolmogorov length ($\Delta x \approx 4 \eta$), the thickness of the laser sheet is an order of magnitude larger ($\approx 50 \eta$). Thus, the velocity field data obtained from our PIV images is rather coarse, and is only used to determine the r.m.s velocity fluctuation $u^\prime$ and the window of interrogation (cf. \S~\ref{sec:window}) wherein the turbulence is approximately homogeneous and isotropic.}

\subsection{Measurement of the flame edge and its fluctuations}

\textcolor{black}{The instantaneous flame edge determined from $\textrm{TiO}_2$ Mie scattering (cf.~figure~\ref{Fig1}c), is described by a curve in parametric form, $(x(s,t),y(s,t))$ where $s$ is the arc-length. (Distinct curves are obtained for the left and right flame edges.) An approximate representation that is more convenient for analysis is given by the explicit curve $x=\xi (y,t)$. These two representations are equivalent except for points where wrinkling causes the flame edge to become a locally multi-valued function of $y$. At such instances, we obtain a single-valued function $\xi$ by considering the leading points of the flame edge, i.e., choosing the point with the smallest value of $x$ for every $y$. This treatment is akin to viewing the flame from the side of the burnt products and making single-point measurements of its surface as it advects past various downstream stations. The curve $x=\xi (y,t)$ thus obtained is termed the leading flame edge (figure \ref{Fig1}d). Such an approach is routinely used in studies of wrinkled flame surfaces \citep{zeldovich1985mathematical, karpov1996test, chterev2018velocity} and simplifies subsequent analysis without altering our key conclusions, as discussed further in \S~\ref{sec:curvature} and Appendix A.} 

To analyse the fluctuation of the flame, which is the primary focus of this work, we first time-average to obtain the V-shaped mean flame edge $\langle \xi (y)\rangle$ (dashed line in figure \ref{Fig1}d). The fluctuations are then defined as $\xi^\prime(y,t) = \xi - \langle \xi \rangle$. \textcolor{black}{Harnessing the transverse ($x$-direction) symmetry of the experimental setup, we combine the measurements of fluctuations obtained from the left and right flame edges to obtain better statistics.}

Detailed information on the experimental facility, measurements, and flame edge detection is provided in the Supplemental Materials.

\vspace{-5pt}
\section{Outer and inner intermittency: two distinct forms of extreme fluctuations}

\begin{figure}
\centering
\includegraphics[width=0.8\textwidth]{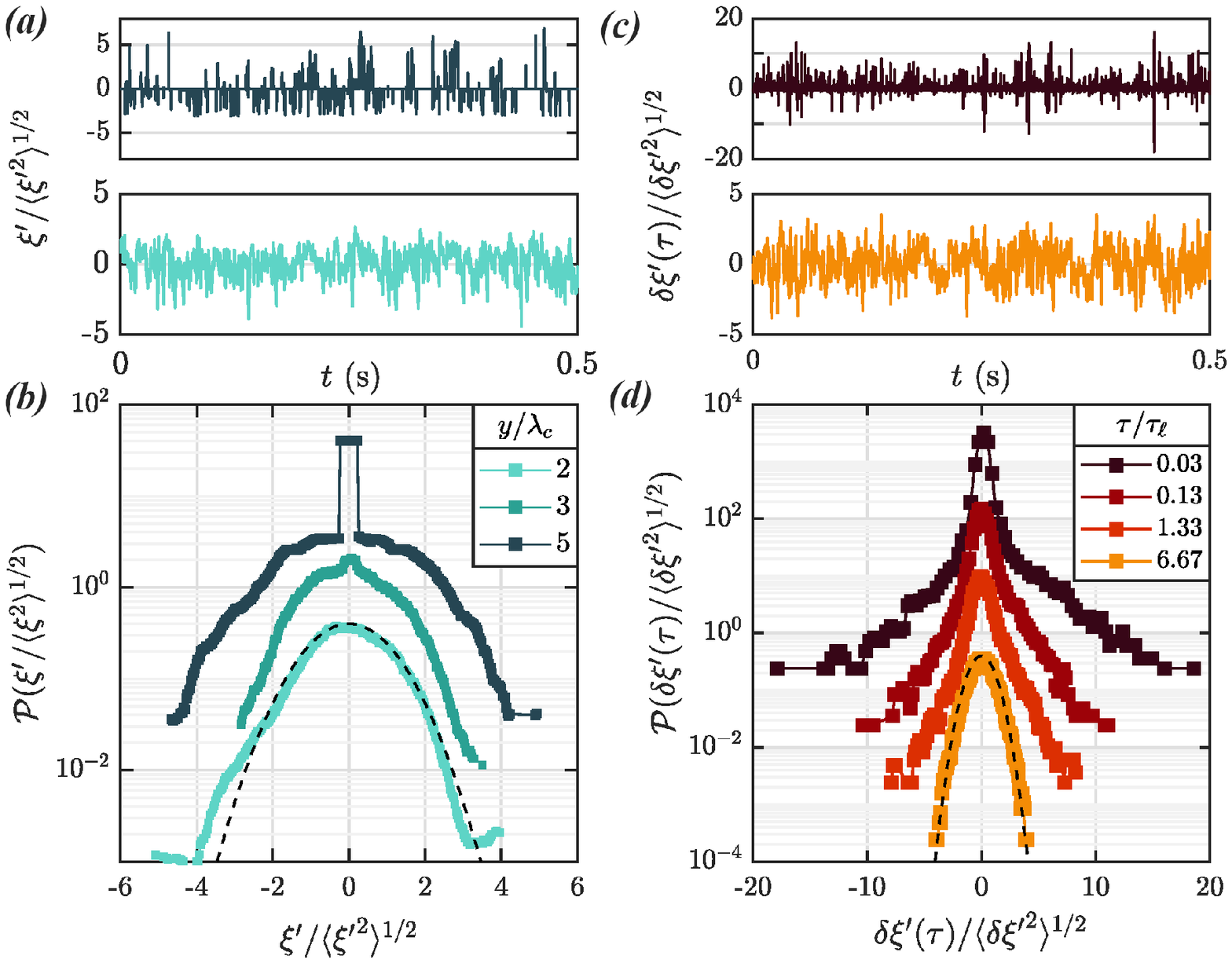}
\caption{\label{Fig2} Distinction between outer and inner intermittency. (a) Time series of the flame fluctuations $\xi^\prime$ normalized by the standard deviation measured at $y=5\lambda_c$ (top) and $y=2\lambda_c$ (bottom). (b) PDFs of $\xi^\prime$ at various axial locations. (c) Time series of increments $\delta\xi^\prime$, normalized by the standard deviation, for $\tau=0.03  \tau_\ell$ (top) and $\tau=6.67\tau_\ell$ (bottom), measured at $y=2\lambda_c$. (d) PDFs of the increment $\delta\xi^\prime$ measured at $y=2\lambda_c$ for various values of the time lag. The data set in all panels corresponds to flame F1. As explained in the text, comparing panels (b) and (d) helps to clearly distinguish large-scale outer intermittency from small-scale inner intermittency. Here, the PDFs have been shifted for clarity and the dashed lines represent $\mathcal{N}(0,1)$ Gaussian fits.}
\end{figure}

Let us begin by considering the fluctuations of the flame at various distances from the flame holder. At relatively large distances, $y/\lambda_c=5$ (where $\lambda_c=\bar{u}_y/f_f$), the flame propagation is erratic and the time series of $\xi^\prime$ (top panel, figure \ref{Fig2}a) exhibits an \textcolor{black}{intermittent behavior. Due to large-scale flapping, the flame undergoes abrupt excursions from the mean (bursts) at some time instances, while failing to propagate to the measurement location at other time instances (off-events).} The off-events are marked by setting $\xi^\prime = 0$, and so the corresponding normalized PDF of $\xi^\prime$ has a sharp peak at $\xi^\prime = 0$ (figure \ref{Fig2}b). Such a PDF has a high kurtosis or flatness factor, $K= \langle {\xi^\prime}^4 \rangle/\langle {\xi^\prime}^2 \rangle^2 =13.51$, and is typical of large-scale outer intermittency. Now, as we move closer to the flame holder, the flame becomes well-maintained and outer intermittency is lost. Indeed, the PDF of $\xi^\prime$ for $y/\lambda_c=2$ (figure \ref{Fig2}b, see also the time series in figure \ref{Fig2}a) has a kurtosis ($K = 3.47$) that is close to the Gaussian value of 3.

This near-Gaussian PDF of $\xi^\prime$, however, veils an intermittency of a different nature, which is revealed by examining the temporal increments of the flame position $\delta\xi^\prime(\tau)=\xi^\prime(t+\tau)-\xi^\prime(t)$. The normalized PDFs of this temporal structure factor, measured at $y/\lambda_c=2$, are presented in figure~\ref{Fig2}(d) for various values of $\tau/\tau_\ell$ ($\tau_\ell = \ell/u^\prime$ is the integral time scale). While the PDF is near-Gaussian for large $\tau$, it develops strongly flared tails for small $\tau$. The kurtosis for $\tau/\tau_\ell = 0.03$ is $K=55.13$, while for $\tau/\tau_\ell = 6.67$, it is $K=3.23$. Comparing the corresponding time series, shown in the bottom and top panels of figure~\ref{Fig2}(c)  respectively, we see that $\delta\xi^\prime(0.03\tau_\ell)$ intermittently undergoes large excursions---tens of times larger than the standard deviation---which are absent in case of $\delta\xi^\prime(6.67\tau_\ell)$.  So, while the large-scale fluctuations at $y/\lambda_c=2$ are non-intermittent, the small-scale fluctuations exhibit extreme-value increments---a clear sign of inner intermittency. 

\begin{figure}
\centering
\includegraphics[width=\textwidth]{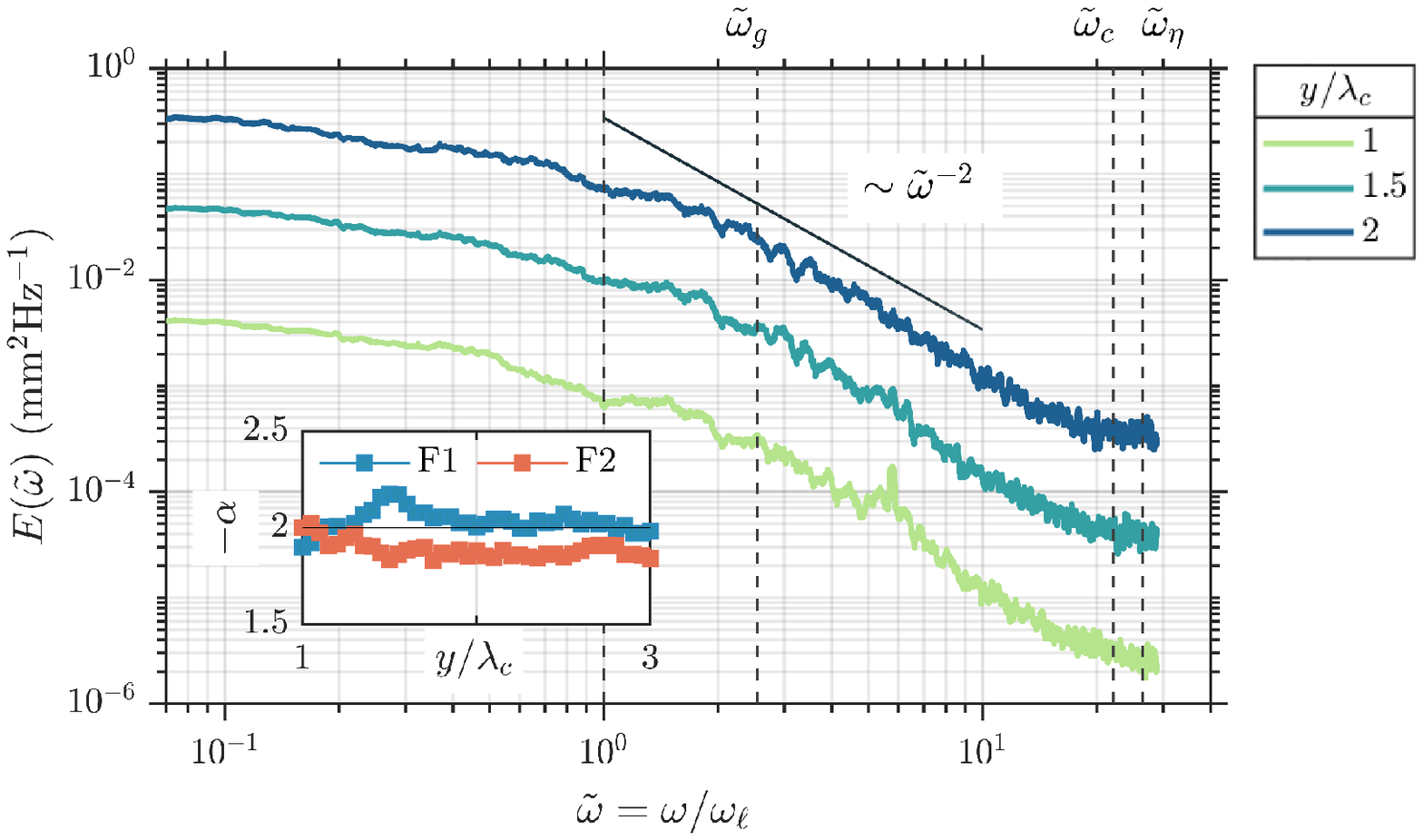}
\caption{\label{Fig3} Temporal power spectral density $E(\tilde{\omega})$ measured at various axial locations $y/\lambda_c$, for flame F1. The power spectra varies as $E(\tilde{\omega})\sim\tilde{\omega}^{-\alpha}$ over an intermediate range of frequencies. The estimated values of the exponent $-\alpha$ at various $y/\lambda_c$ are shown in the inset for both flames F1 and F2. In both cases, the exponent is close to -2; the corresponding scaling behavior is depicted by the solid line in the main panel. The frequency corresponding to Gibson ($\tilde{\omega}_g$), Corrsin ($\tilde{\omega}_c$) and Kolmogorov ($\tilde{\omega}_\eta$) scales have been indicated by dashed lines.}
\end{figure}

\section{Power-law scaling of the frequency spectra}

Before characterizing this intermittency further, it is instructive to examine the power spectra of $\xi^\prime$ in frequency space, which is closely related to the variation of the the second moment of $\delta\xi^\prime$ with $\tau$. The frequency spectra is presented in figure \ref{Fig3} for three axial locations. At $y/\lambda_c=1$, we see a minor imprint of the external vibration of the flame holder, in the form of a small peak at the forcing frequency $\omega_f/\omega_{\ell} = 3.75$, where $\omega_{\ell}=2 \pi/\tau_{\ell}$. For $y/\lambda_c=1.5$ and 2, there is no trace of the external forcing, and the flame's fluctuations are dominated by its response to the turbulent flow. Interestingly, for frequencies beyond $\omega_{\ell}$, a self-similar power-law is seen to emerge.

To understand the origin of this power-law, let us begin with the flame fluctuation spectra in spatial wavenumber ($k$) space, which has been well studied. Indeed, for a flame of finite thickness susceptible to diffusive effects [$Da \sim \mathcal{O}(1)$], in an isotropic and homogeneous turbulent flow, Kolmogorov's phenomenology \citep{peters1992spectral,chaudhuri2011spectral}  leads to a spectra with a power-law behaviour, 
\begin{equation}
    \Gamma(k) \sim k^{-5/3},
    \label{eq:kspectra}
\end{equation} 
in the subrange $k_{\ell}<k<k_{c}$. Here $k_{\ell} = 2 \pi/\ell$ corresponds to the large integral scale while $k_{c}=2 \pi/\eta_c$ is the wavenumber of the Corrsin length scale $\eta_c=Sc^{-3/4}\eta$ ($Sc=\nu/\mathcal{D}_M$, where $\mathcal{D}_M$ is the Markstein diffusivity) after which diffusive effects within the flame become dominant. (The role of kinematic restoration is discussed below.) This subrange lies within the inertial range of the turbulent flow, $k_{\ell}<k<k_{\eta}$, where $k_{\eta}=2 \pi/\eta$ corresponds to the viscous Kolmogorov length $\eta$ ($k_{\eta} > k_c$ as $Sc < 1$ for our flame). Notably, the scaling in (\ref{eq:kspectra}) is the same as that for a passive scalar in the inertial-convective range \citep{oboukhov1949structure, corrsin1951spectrum, davidson2015turbulence}, which is consistent with the fact that effects due to the flame's propagation do not play a role over this range of scales. In summary, the inertial range velocity fluctuations educe an apparently self-similar response from the flame surface, which however is cut-off by large-scale effects for $k < k_{\ell}$ and by diffusion within the flame for  $k > k_c$.

Now, in order to translate this picture to the frequency domain, we assume that flame fluctuations with wavenumber $k$ are most strongly influenced by a turbulent eddy of size $2 \pi/k$, whose typical velocity according to inertial range scaling is $u^\prime \sim k^{-1/3}$. We can then relate the wavenumber of flame fluctuations to the frequency as $\omega \sim ku^\prime \sim k^{2/3}$. Then, using (\ref{eq:kspectra}) and the fact that the frequency spectra $E(\omega)$ must satisfy $\int E({\omega})d{\omega}=\int\Gamma({k})d{k}$, we obtain:
\begin{equation}
E(\omega) \sim {\omega}^{-2}.
\label{Eq-3}
\end{equation}

The exponent of the power-law subrange of the frequency spectra is presented in the inset of figure~\ref{Fig3}, and is seen to be close to this prediction of $-2$ for a range of axial locations, $\lambda_c<y<3\lambda_c$. This figure also shows the frequencies $\omega_{\ell}$, $\omega_{c}$ and $\omega_{\eta}$ corresponding to the length scales $\ell$, ${\eta}_c$ and $\eta$, respectively. The power-law is seen to commence after $\omega_{\ell}$, as expected, and then carry on for about a decade. However, the temporal resolution of our measurements is insufficient to resolve the diffusive cutoff and subsequent stretched-exponential decay of the spectra beyond $\omega_c$ \citep{peters1992spectral,chaudhuri2011spectral}. 

Figure~\ref{Fig3} also shows the frequency $\omega_g$, which corresponds to the Gibson length $\ell_g=(s_L/v^\prime)^3\ell$, at which the flame propagation speed becomes comparable to the turbulent velocity fluctuations. This scale could potentially cut off the power-law scaling due to kinematic restoration effects that act to smooth out turbulence-induced flame fluctuations \citep{lieuwen2021unsteady}. However, previous work indicates that for flames of finite thickness, this effect plays a minor role, possibly being counter-acted by thermal expansion effects, so that the power-law behavior persists until the Corrsin scale ($\omega_c$), after which it is terminated by diffusive effects within the flame \citep{peters2000modification, gulder2000flame, shim2011local, chatakonda2013fractal}.

\section{Anomalous scaling of temporal increments}
Let us now return to the issue of inner intermittency and its characterization. This is best done by examining the scaling of the structure functions $S_p(\tau)$, defined as the $p^{\rm th}$ moment of the increment $\delta\xi^\prime(\tau)$ \citep{frisch1995turbulence,falcon2007observation}:
\begin{equation}
S_p(\tau;y) \equiv \langle[\delta\xi^\prime(\tau;y)]^p\rangle=\langle[\xi^\prime(t+\tau;y)-\xi^\prime(t;y)]^p\rangle.
\label{Eq-4}
\end{equation}
For the second-order structure function, which can be determined entirely from the power spectra \citep{davidson2015turbulence}, we have $S_2 \sim (2\pi/\tau) E(2\pi/\tau) \sim \tau$. For the $p^{\rm th}$ moment then, a naive expectation would be $S_p = \langle{[\delta \xi^\prime]}^p\rangle\sim \tau^{p/2}$; this would be true if the flame fluctuations were non-intermittent and perfectly self-similar. However the presence of extreme-value increments, evident in the flared-tail PDFs of figure~\ref{Fig2}(d), causes the higher-order moments to have increasingly large values as $\tau$ decreases. So, for intermittent fluctuations, we expect $S_p \sim \tau^{\zeta_p}$ with $\zeta_p$ becoming increasingly smaller than $p/2$ as $p$ increases. 

\begin{figure}
\centering
\includegraphics[width=\textwidth]{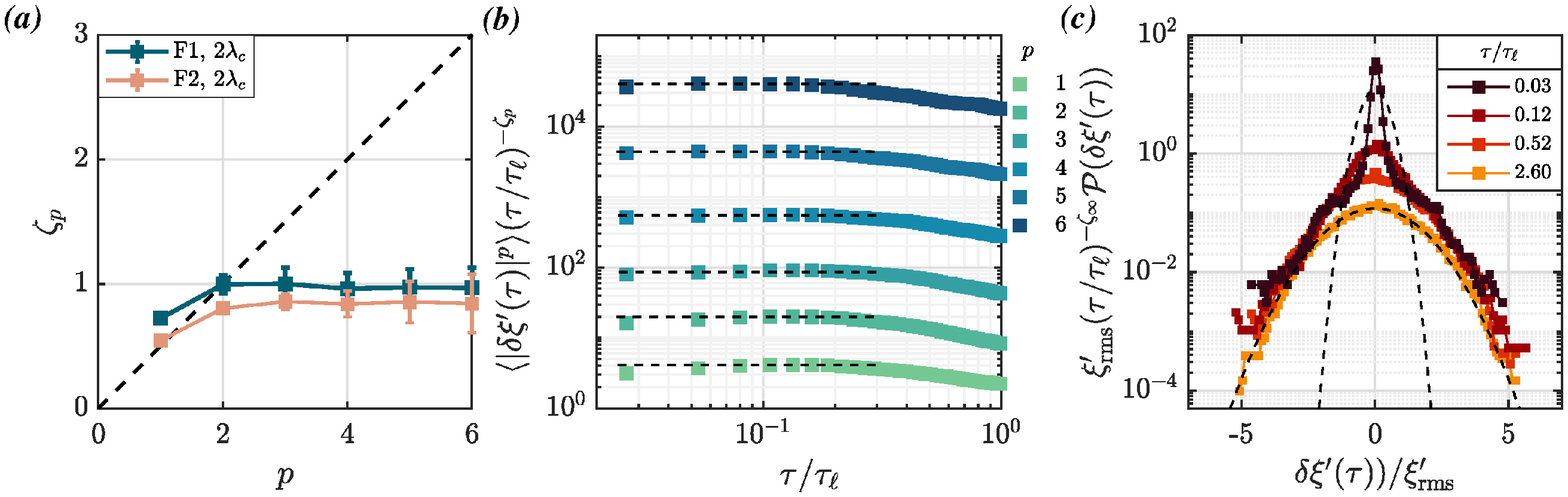}
\caption{\label{Fig4} (a) The variation of the scaling exponents $\zeta_p$ with the order $p$ of the structure function, for both flames at $y/\lambda_c=2$. The dashed line indicates the non-intermittent limit of $\zeta_p=p/2$. The strong deviation of $\zeta_p$ from this limit for $p>2$, implies that the exponents scale anomalously and that the flame fluctuations are strongly intermittent. The error bars represent the standard deviation of the measured values obtained from different time series (see \S~S6 in Supplemental Materials). (b) Plots of the structure functions upto order $p=6$, compensated by the estimated scaling $\tau^{-\zeta_p}$, which illustrates their scaling behaviour (for flame F1 measured at $y=2\lambda_c$). (c) PDFs of $\delta{\xi^\prime}(\tau)$, for various values of $\tau$, multipled with $\xi^\prime_{\rm rms}(\tau/\tau_{\ell})^{-\zeta_{\infty}}$, where $\zeta_{\infty}$ is the saturated value of $\zeta_{p}$ estimated from panel (a). The dashed lines are Gaussian fits for $\tau/\tau_\ell=0.03$ and $2.60$. The collapse of the tails of the PDFs is striking, especially when compared with the behaviour of the Gaussian fits.}
\end{figure}

This is exactly what we observe in figure~\ref{Fig4}(a), which presents the values of $\zeta_p$ as a function of $p$, up to the sixth order, for both flame configurations, at $y = 2\lambda_c$. Figure~\ref{Fig4}(b) illustrates the corresponding power-law scaling of $S_p$ for $\tau < \tau_{\ell}$. Equivalent results are obtained at other axial locations in the interval $\lambda_c<y<3\lambda_c$, wherein the spectra exhibited a power-law exponent close to $-2$ (cf. Fig~\ref{Fig3}). We also estimated $\zeta_p$ using the procedure of extended self-similarity \citep{benzi1993extended}, which takes advantage of the fact that $S_p/S_2$ scales as $\tau^{\zeta_p/1}$ over an extended range of $\tau$, and obtained values nearly identical to those shown in figure~\ref{Fig4}(a). 
The dramatic departure of $\zeta_p$ from $p/2$, for $p$ beyond second-order, makes evident the intensely intermittent nature of the small-scale fluctuations of the flame.  

The saturation of the $\zeta_p$ exponents with increasing $p$, seen in figure \ref{Fig4}(a), is reminiscent of the anomalous scaling behavior of passive scalar turbulence \citep{celani2000universality, celani2001fronts, watanabe2006intermittency}, wherein the saturation arises due to steep ramp-cliff structures in the concentration field. 
The width of these structures decreases as the diffusivity is reduced, yet the amplitude of the concentration jump remains near the root-mean-square value of concentration fluctuations \citep{celani2001fronts}.
In our case, figure \ref{Fig4}(a) implies that as the diffusive Corrsin scale is reduced (keeping $Da$ and $Sc$ constant), the extreme-valued flame fluctuations would undergo a displacement of the order of the root-mean-square value, $\xi^\prime_{\rm rms}=\langle{\xi^\prime}^2\rangle^{1/2}$, in an ever-shortening time. To see this, note that in the limit of large $p$, $S_p^{1/p} \sim \langle\delta{\xi^\prime(\tau_{\ell})}^p\rangle^{1/p} (\tau/\tau_{\ell})^{(\zeta_p/p)}$ is an estimate of the magnitude of extreme increments. So, given that $\zeta_p$ saturates as $p$ increases, we see that the magnitude of extreme increments does not decrease with $\tau$, but rather remains comparable to $\langle\delta{\xi^\prime(\tau_{\ell})}^p\rangle^{1/p} \sim \langle{\xi^\prime}^p\rangle^{1/p} \sim \xi^\prime_{\rm rms}$ [$\xi^\prime$ has an approximately Gaussian distribution as seen in Fig.~\ref{Fig2}(b)]. This behaviour is illustrated in Fig.~\ref{Fig4}(c), wherein the tails of the PDFs of $\delta{\xi^\prime}$, for various values of $\tau$, are seen to have the same width, and in fact collapse when the PDFs are multiplied with $\xi^\prime_{\rm rms}(\tau/\tau_{\ell})^{-\zeta_{\infty}}$ [$\zeta_{\infty}$ is the saturated value estimated from Fig.~\ref{Fig4}(a)]. 

\section{Origin of extreme-valued temporal increments}
\label{sec:curvature}

The extreme temporal increments implied by the saturating exponents in figure~\ref{Fig4}(a) have two possible causes: (i) rapid fluctuation events in which the flame advances and retreats very quickly, or (ii) advection of coherent spatial structures on the flame surface, such as cusps, past a fixed measurement location. The second scenario has been proposed as the cause of intermittency in temporal fluctuations of a free-surface exhibiting gravity-capillary wave turbulence \citep{falcon2007observation}. For flames, and propagating surfaces in general, cusp-like features are typical \citep{Law2000,Yang2017} and indeed appear quite frequently on the flame edge in our study (see figure~\ref{Fig1}c). Near such points, the flame edge typically traces out a large excursion in the transverse $x$ direction over a short distance in the longitudinal $y$ direction. So, when such a cusp-like structure is advected past the measurement location ($y=2\lambda_c$ in figure~\ref{Fig4}) by the mean flow, it will register as an extreme valued increment of the flame position. The time-scale of such events is estimated in \S~\ref{sec:resol} to be approximately $10^{-4}$ s, which is strikingly similar to the time-scale of the interval $\tau$ at which the PDF of $\delta\xi^\prime(\tau)$ begins to develop strongly flared tails [see figure~\ref{Fig2}(d) wherein $\tau = 0.03 \,\tau_\ell \approx 10^{-4}$ s].  Indeed, an examination of the temporal evolution of the flame surface in tandem with the time-trace of the temporal increment (presented in Movie M1 in the Supplemental Materials) strongly suggests that this scenario predominates and that the anomalous scaling of figure~\ref{Fig4}(a) is due to the advection of cusp-like structures along the flame edge.

\begin{figure}
    \centering
    \includegraphics[width=\textwidth]{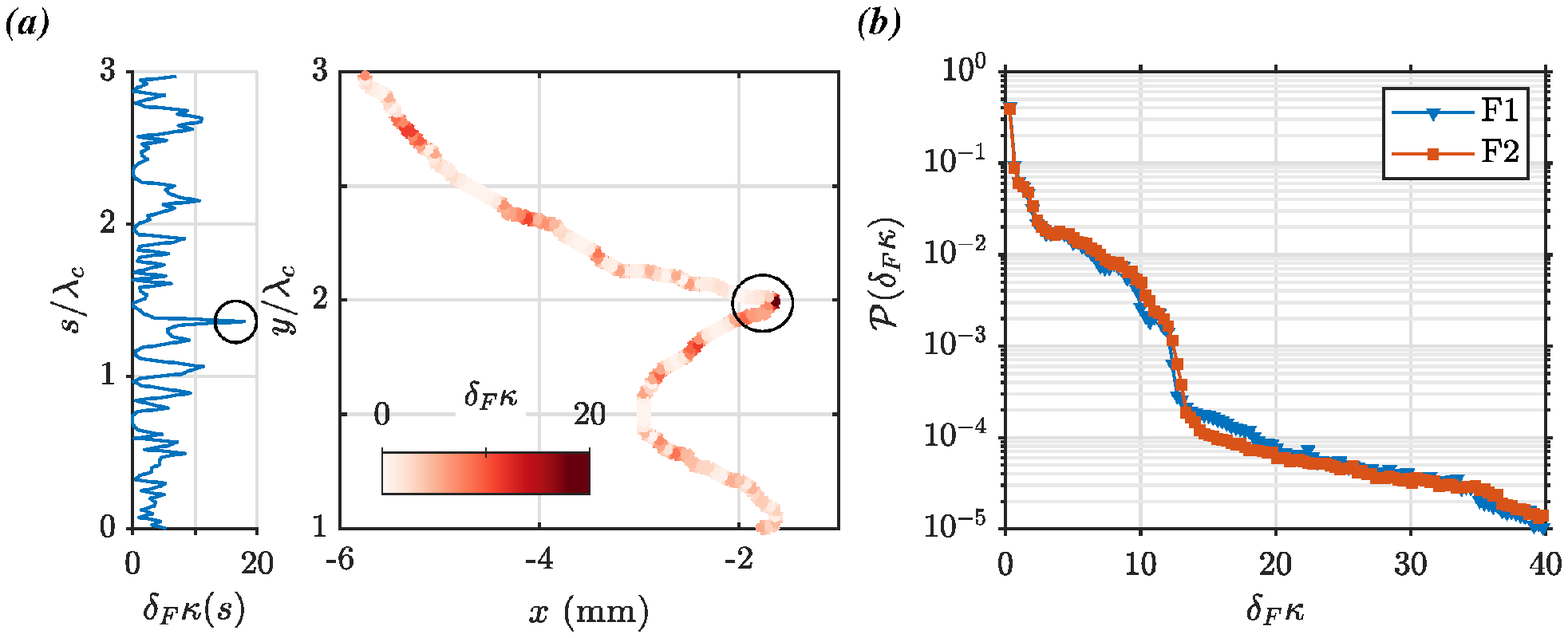}
    \caption{\textcolor{black}{(a) Illustration of the variation of the curvature $\kappa$ along the flame surface in the presence of a cusp-like structure. The curvature attains a large value at the cusp (right panel) and thus appears as a prominent spike in the plot of $\kappa$ as a function of the arc-length $s$ (left panel). Note that the curvature has been nondimensionalized using the flame thickness $\delta_F$. (b) Stationary PDF of the curvature $\kappa$, shown for flames F1 and F2, exhibiting a flared positive-tail which reflects the presence of cusp-like structures on the flame surface. The PDFs are constructed by calculating the curvature along the flame edge contained within the window of interrogation (\S~\ref{sec:window}) and over time.}}
    \label{Fig5}
\end{figure}

As a quantitative check, we now calculate the curvature of the flame edge and examine whether extreme values of the temporal increment $\delta{\xi^\prime}$ occur simultaneously with extreme values of curvature, which would correspond to cusps. 
Using the parametric representation of the flame edge $(x(s,t),y(s,t))$, where $s$ is the arc-length, the curvature $\kappa$ is calculated as follows \citep{ArisVectors}: 
\begin{equation}
\label{eq:curvature}
    \kappa(s,t) = \left((\partial_{ss}x)^2+(\partial_{ss}y)^2\right)^{1/2}.
\end{equation}
We construct the curve $(x(s,t),y(s,t))$ using fourth-order spline interpolation based on points spaced equally along the flame edge, such that the inter-point distance ($ds=\sqrt{dx^2+dy^2}=0.1$ mm) is greater than the pixel size $\Delta x$. This allows us to evaluate the derivatives in (\ref{eq:curvature}) and obtain the curvature as a function of the arc-length $s$ \citep[see][for a similar calculation for material loops in turbulence]{bentkamp2022statistical}. Figure \ref{Fig5}(a) illustrates the typical variation of the curvature along the flame edge when it has a cusp-like feature: we see that the cusp corresponds to a spike in the curvature profile. Figure~\ref{Fig5}(b) presents the stationary PDF of curvature values sampled by the flame edge, within the interrogation window (box in figure~\ref{Fig1}d) and over time. The heavy-tail of the distribution is a consequence of the extreme curvature values associated with cusp-like features of the flame surface.

\begin{figure}
    \centering
    \includegraphics[width=\textwidth]{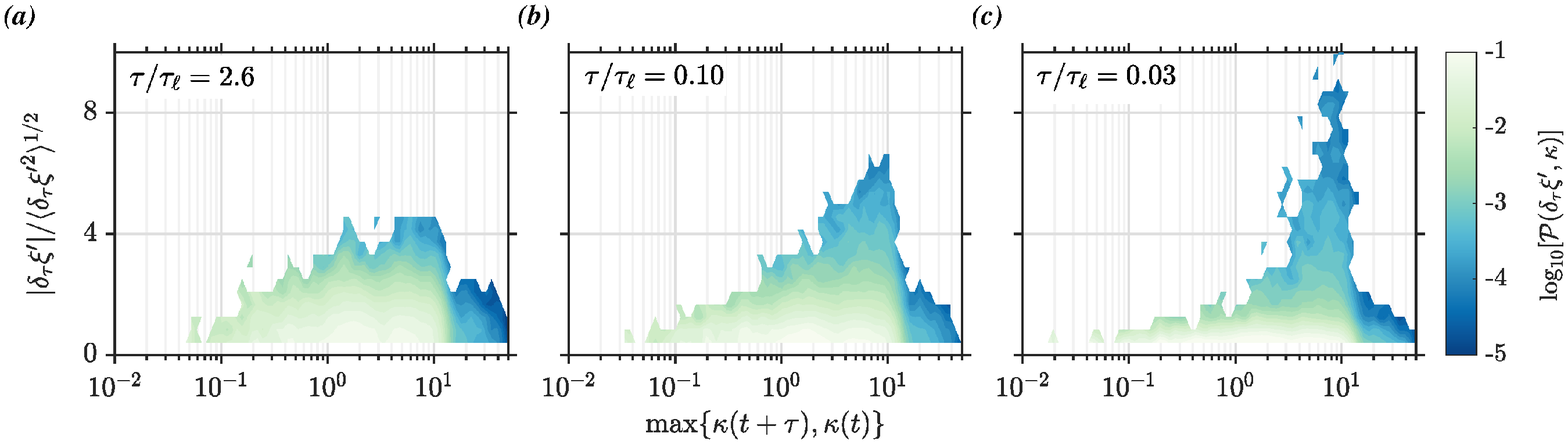}
    \caption{Pseudo-color plot of the joint probability distribution function of the increment $\delta\xi^\prime (\tau;y)$ and the maximum of the flame curvature values at times instances $t$ and $t+\tau$, as measured at $y = 2 \lambda_c$. The three panels correspond to different values of the time increment $\tau/\tau_\ell$:  (a) 2.67, (b) 0.10 and (b) 0.03. The extreme-valued increments which appear as $\tau$ decreases show a clear correlation with extreme values of the curvature. These results correspond to flame F1 and are measured at $y=2\lambda_c$; similar results are obtained for flame F2 and for other measurement locations within $1<y/\lambda_c<3$. The colored contours correspond to the the logarithm of the joint PDF.} 
    \label{Fig6}
\end{figure}

Having calculated the flame curvature, we next examine its correlation with the temporal increment of the flame fluctuations. Each measurement of the increment $\delta\xi^\prime(\tau;y^*)$ is associated with curvature values at two time instances, $\kappa(t;s_1)$ and $\kappa(t+\tau;s_2)$, where the values of $s_1$ and $s_2$ are such that $y(s_1)=y(s_2)=y^*$ corresponds to the measurement location. Note that there will be more than one value of $s_1$, or $s_2$, when the flame edge becomes locally multi-valued. We consider the maximum of these curvature values $\max\{{\kappa}(t),{\kappa}(t+\tau)\}$ and depict its joint probability distribution function with the magnitude of the associated increment of the flame position $|\delta\xi^\prime(\tau)|$ in figure~\ref{Fig6}, for various time intervals $\tau$. Clearly, the extreme-values of the temporal increment which arise as $\tau$ decreases are strongly correlated with extreme-values of the flame curvature. This is entirely consistent with our observation that large and rapid temporal fluctuations of the flame position are registered when cusp-like structures are advected past the measurement location.

Cusp-like features and wrinkling in general can cause the flame edge to become a locally multi-valued function of $y$. This fact is accounted for in the calculation of curvature $\kappa(s,t)$ which is based on the arc-length parameterization, but not in the measurement of increments $\delta\xi^\prime$ which is based on the single-valued leading edge $\xi(y,t)$. This raises the concern that multi-valued regions of the flame edge may appear as artificially abrupt variations in $\xi(y,t)$ which would in turn produce spurious large values of the increment. In Appendix \ref{sec:ApdxA}, we carry out two checks which show that our detection of inner intermittency is not an artifact of the treatment of the flame edge. First, we modify how we obtain the single-valued function $\xi(y,t)$: instead of taking the points closest to the $y$ axis, we now locally average the flame edge in regions where it is multi-valued. This procedure would smooth out artificially abrupt variations in $\xi(y,t)$. Second, we retain the leading edge profile, but eliminate all data points where the flame edge is multi-valued, i.e., we only calculate increments when $\xi(y)$ is single valued at both times, $t$ and $t+\tau$. For both cases, we find that anomalous scaling persists, which shows that the inner intermittency detected in this study is a genuine feature of the flame dynamics.

\section{Concluding Remarks}
\label{sec:conclusion}
To summarize, we have seen that a well-maintained flame surface, devoid of large-scale bursts associated with outer-intermittency, contains a subrange of scales wherein the flame surface merely responds to the fluctuations of the incident turbulent flow and exhibits power-law scaling that is entirely determined by the inertial-range scaling of the flow. However, this apparent simplicity comes along with a caveat, which is the key result of our work: the flame fluctuations are intensely intermittent with structure functions that exhibit strongly anomalous scaling. \textcolor{black}{The associated extreme events, which originate from cusp-like structures that are advected along the flame edge, have important implications for the modeling of turbulent premixed flames. For example, cusps with their extremely large values of flame curvature will affect the mean turbulent flame speed \citep{law2010combustion,humphrey2018premixed,dave2020evolution}}. Furthermore, closure models of turbulent flame speeds and volumetric heat release rate, which depend on the fractal dimension of the flame surface, assume perfect self-similarity and so may be improved by accounting for inner intermittency 
\citep{charlette2002power, gulder2007contribution, roy2020fractal}. 

\textcolor{black}{It is intriguing to consider how the cusp-like features on the flame surface are related to the small-scale structures, such as ramp-cliffs, of the underlying scalar fields. While ramp-cliff structures have been observed in DNS simulations of premixed combustion \citep{wang2007experimental, cai2009investigation}, their connection to intermittent fluctuations of the flame surface is unknown. More generally, the question of how the statistics of the flame surface are connected to that of the reacting scalar fields of combustion, possibly via a flame indicator function~\citep{thiesset2016geometrical}, deserves further study especially in light of the extensive literature on the turbulent transport of conserved scalars~\citep{warhaft2000passive,Falkovich2006lessons}; one would require simultaneous high-resolution measurements of the flame surface and the scalar fields, which while beyond our current scope is an important task for future work.}

It is also interesting to note that the equation for the propagation of a thin premixed flame resembles the Kardar-Parisi-Zhang (KPZ) equation \citep{kerstein1988field, KPZ} [in turn closely related to the Burgers equation~\citep{BecBurg2007}], whose dynamics in the presence of additive noise is well-studied \citep{Verma2000}. However, a crucial difference arises due to the advection of the flame by the turbulent flow, which, if modeled as a random flow, appears as multiplicative noise. This results in fundamentally distinct dynamics \citep{Yakhot1988,kerstein1988field}, for example, the propagating flame attains a statistically stationary mean speed in contrast to the power-law growth predicted by the KPZ equation with additive noise. Thus, in light of the present results, it would be interesting to explore the intermittent properties of the KPZ equation with multiplicative and spatio-temporally correlated noise.

Finally, while we have focused on temporal measurements, the cusp-like structures of the flame surface will certainly give rise to inner intermittency in space, so that spatial structure functions obtained from a flame edge profile at a single snapshot in time should scale anomalously. Establishing this experimentally would require a very large flame in order to capture the required range of spatial scales, a challenge that will hopefully be taken up in the future. Other important questions raised by our work include how the inner intermittency of the self-similar range relates to the extreme-value statistics of the sub-diffusive dissipative scales \citep{hamlington2012intermittency, chaudhuri2017flame}, and whether the scaling exponents for a premixed flame are universal, as they seem to be for a passive scalar \citep{watanabe2006intermittency}.

\section*{Acknowledgements}
This research was conceptualized following a visit to the International Centre for Theoretical Sciences (ICTS), India for participating in the program - Turbulence: From Angstroms to light years (Code: ICTS/Prog-taly2018/01). The authors thank Luke Humphrey (Georgia Tech) for sharing his experimental data. The authors benefited from discussions at the Inter Group Meetings held at IIT Madras, ICTS Bangalore, and IIT Bombay. A.R. and J.R.P. also thank Samriddhi Sankar Ray (ICTS) and Jeremie Bec (MINES ParisTech) for insightful discussions and comments.

A.R. is grateful for the HTRA Ph.D. fellowship from MHRD, India. J.R.P. is thankful for financial support from the IIT Bombay IRCC Seed Grant and from the DST-SERB grant (SRG/2021/001185). T.C.L. gratefully acknowledges the support received from Air Force Office of Scientific Research (Contract no. FA 9550-20-1-0215), contract monitor Dr. Chiping Li. R.I.S. gratefully acknowledges funding from the Institute of Excellence Grant (SB/2021/0845/AE/MHRD/002696) and the J. C. Bose Fellowship (No. JCB/2018/000034/SSC).

\appendix

\begin{figure}
\centering
\includegraphics[width=.8\textwidth]{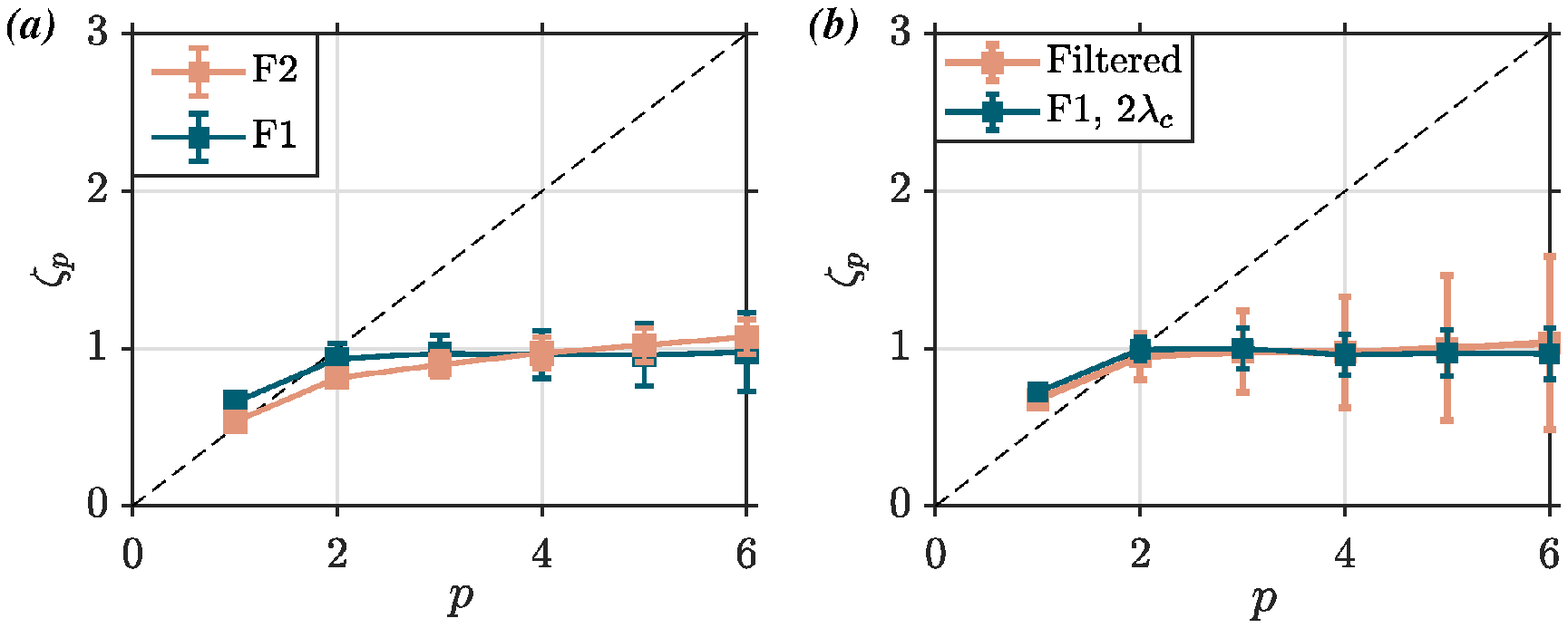}
\caption{{\color{black} Anomalous scaling of the exponents of the structure function, calculated after implementing two procedures that are designed to reduce or eliminate any artificial large increments in the flame position that arise from multi-valued folds or wrinkles being described by the single-valued leading flame edge. First, for panel (a), we use the locally averaged flame edge profile, which averages over multi-valued regions of the flame thereby providing a much smoother representation of multi-valued folds when compared to the leading edge. Second, for panel (b), we use a filtered data set of increments of the leading edge, obtained by removing any increment value that is calculated using time instants when the flame is multi-valued. The filtered data set is therefore entirely free from the influence of multi-valued regions of the flame, and the exponents of the corresponding structure functions are compared with those of the full data set, for the case of flame F1. Clearly, the anomalous scaling behavior persists for the locally averaged flame edge, as well as for the filtered data (which however has enlarged error bars due to a reduction of data points).}}
\label{Fig:Apdx1}
\end{figure}

{\color{black}
\section{Anomalous scaling persists after smoothing or eliminating multi-valued flame wrinkles}\label{sec:ApdxA}

Wrinkling of the flame in turbulence causes the flame edge $(x(s),y(s))$ to become locally multi-valued, so that there will be multiple values of $x(s)$ for each $y(s)$. In the main text, such multi-valued regions of the flame edge are converted to a single-valued function of $y$---the leading flame edge---by choosing the value of $x$ with the smallest magnitude. While this treatment allows for a straightforward definition of the flame fluctuation and its temporal increment, it does introduce the possibility of multi-valued folds in the flame edge $(x(s),y(s))$ being registered as artificially abrupt variations in the leading flame edge $\xi(y)$. This could in turn produce artificial large increments when these folds are advected past a fixed longitudinal measurement location. Here, we carry out two checks which are designed to reduce or eliminate the effect of such multi-valued folds on the statistics of the flame increment; if the anomalous scaling behaviour persists then we can be confident that it is a genuine feature of the flame dynamics.

First, rather than using the leading flame edge, we construct a locally averaged flame edge $\tilde{\xi}(y,t)$, where for each value of $y$ we set $\tilde{\xi}$ to be equal to the average of all the corresponding values of $x$ (mathematical definitions for both flame edges is given in the Supplemental Materials). Any multi-valued folds which appear as abrupt variations in $\xi(y,t)$ will be significantly smoothed out in $\tilde{\xi}(y,t)$. Of course, where the flame is single valued, ${\xi}$ and $\tilde{\xi}$ will be identical. We then define the fluctuations and the increment as $\tilde{\xi}^\prime(y,t) = \tilde{\xi} - \langle \tilde{\xi} \rangle$ and $\delta\tilde{\xi}^\prime(\tau)=\tilde{\xi}^\prime(t+\tau)-\tilde{\xi}^\prime(t)$ respectively, and determine the scaling exponents $\zeta_p$ just as we do in the main text. The results, presented in figure~\ref{Fig:Apdx1}(a), show strongly intermittent behaviour similar to that seen in figure~\ref{Fig4}(a). 

As a second check, we return to the leading flame edge, but now entirely eliminate any points where the flame edge is multi-valued from the calculation of the scaling exponents. For this, we first flag any time instant at which the flame edge $(x(s),y(s))$ is multi-valued. Then, while calculating the increments $\delta{\xi}^\prime=\xi^\prime(t+\tau)-\xi^\prime(t)$, we check to see if either of the two data points are flagged (i.e. if the flame edge is multi-valued at either time instant), and if so we discard the corresponding increment value. The filtered set of data values thus obtained correspond to increments $\delta{\xi}^\prime(\tau)$ such that the flame edge is single-valued at both $t$ and $t+\tau$. This procedure will entirely elimininate any large-valued increments due to folds in the flame edge being advected past the measurement location $y$. The scaling exponents obtained from this filtered data set are compared with those obtained from the unfiltered data in figure~\ref{Fig:Apdx1}(b). We see that the strongly anomalous scaling behaviour persists, although the loss of data due to filtering enlarges the error bars especially for large values of the order $p$.

Taken together, these two checks demonstrate that the inner intermittency of the flame fluctuations detected in this work is \textit{not} an artifact of representing the edge by a single-valued function. As shown in \S~\ref{sec:curvature}, the extreme-valued increments arise from the advection of cusp-like structures past the measurement location. These cusp-like features correspond to genuine sharp variations of the flame edge, and so representing them by the single-valued leading edge does not introduce artificial abrupt variations even when the cusps are multi-valued. So, though the precise values of the flame fluctuation and its increment depends on the manner in which the flame edge is described, their statistical features including anomalous scaling are qualitatively alike.
}

\bibliographystyle{jfm}
\bibliography{references}

\end{document}


\maketitle

\section{Experiments, instrumentation and diagnostics}
\subsection{Setup}
\begin{figure*}
\centering
\includegraphics[width=\textwidth]{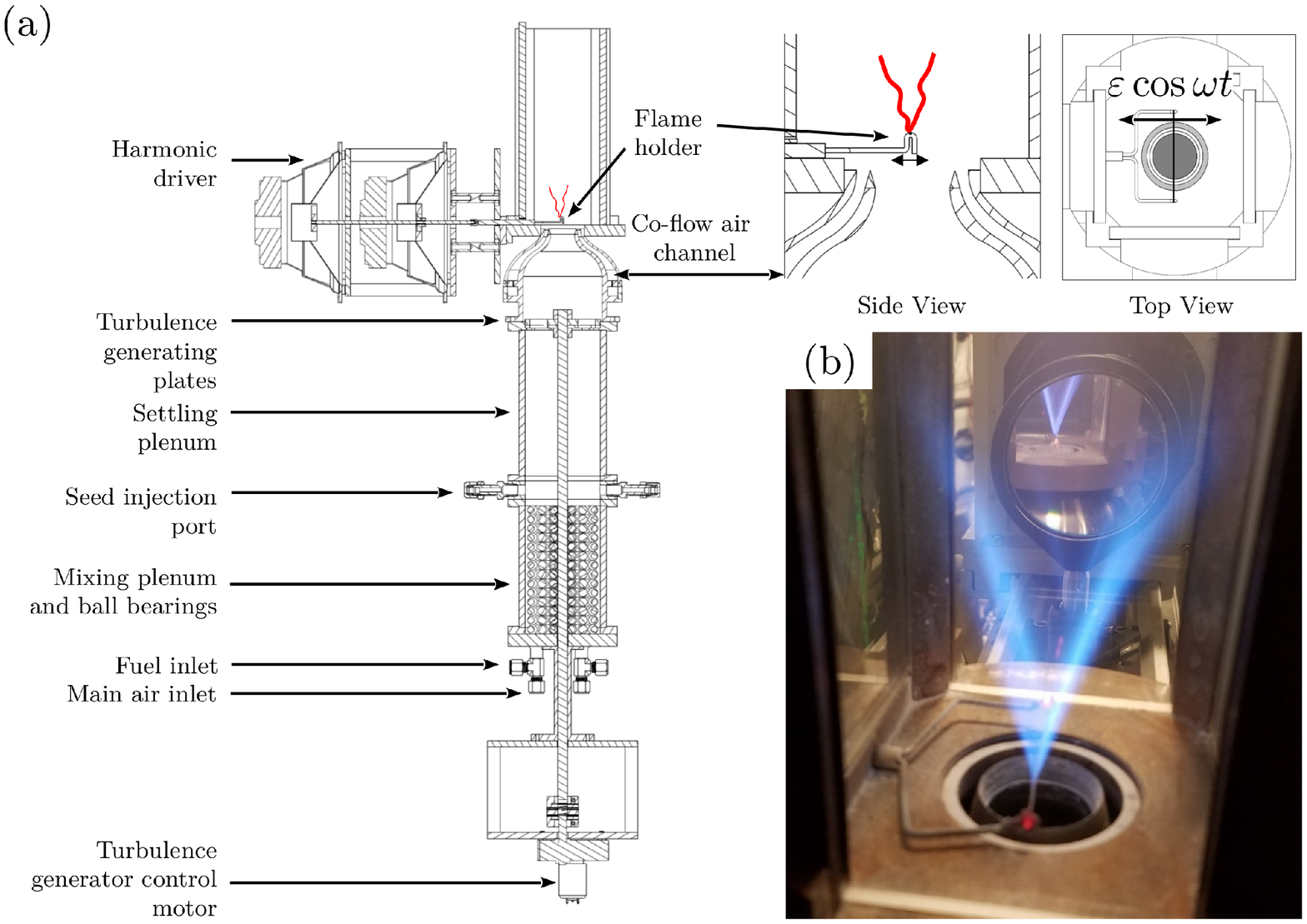}
\caption{\label{FigS1_ExptSetup} Turbulent V-flame facility. (a) Schematic of the combustor setup. (b) Illustrative photograph of the V-flame. Adapted from \citep{humphrey2018premixed} with permission from Cambridge University Press.}
\end{figure*}

The turbulent combustor used in our study of small-scale intermittency is shown in Fig. \ref{FigS1_ExptSetup}. This experimental configuration was designed to assess how flame dynamics are affected by broadband forcing due to turbulence in addition to narrowband forcing due to an oscillating flame-holder \citep{petersen1961stability, kornilov2007experimental, truffaut1999experimental}.

Air and methane ($\text{CH}_4$) enter the premixing chamber through a port at the bottom. The premixing chamber is packed with ball bearings to facilitate thorough mixing of the fuel and air. Next, the mixture of air and $\text{CH}_4$ enters the settling chamber and then passes through the turbulence generator and onward to the combustion chamber through a nozzle. The entry into the combustion chamber is aided by a co-flow of air, injected through the co-flow air channel at the bottom of the nozzle. The main nozzle has an exit diameter of $27.4$ mm. The co-flow is velocity matched to the main flow and ensues out of an annulus with an outer diameter of $36.3$ mm. The main air and fuel flow supply are controlled using Aalborg GFC-$67$, $0-500$ L/min and Omega FMA-$5428$, $0-50$ L/min mass flow controllers, respectively. The co-flow is controlled using an Omega FMA-$1843$ gas flow meter and needle valve. All the mass flow controllers have an uncertainty of $\pm1\%$. The maximum uncertainty in the reported values of equivalence ratio ($\phi$), velocity ($\bm{v}$) and Reynolds number ($Re$) are $\pm2\%$, $\pm1\%$ and $\pm1\%$, respectively.

An electrically heated flame holder ignites the flame. The flame holder is a nichrome wire ($0.81$ mm, $20$ American Wire Gauge) and is heated by $6-12$ V alternating current and held 10 mm above the exit plane of the main nozzle. The flame holder is oscillated, transverse to the oncoming jet flow, at different frequencies and with different forcing amplitudes, using two modified $90$ W Goldwood speakers connected in parallel. The input signal to the speakers is generated using a function generator and amplified by two linear amplifiers. 
Experiments reported in this paper were conducted for a forcing frequency of $f_f = 1250$ Hz and amplitude of $\langle\varepsilon(f_f)\rangle\approx 0.26$ mm. The amplitude of forcing is determined from the power spectrum of the measured time series of the position of the flame holder.

The turbulence generator consists of two plates with several pie-shaped slots cut through them. The bottom plate is fixed (stator), and the top plate (rotor) can be rotated over a $28\pm0.25^\circ$ range. By rotating the rotor, it is possible to change the blockage ratio from $69$\% to $97$\%, which in turn enables us to vary the turbulent intensity, $v^\prime/\bar{v}_y$, in the range of $8$\% to $36\%$. Under isothermal conditions, the turbulent flow so generated exhibits a Kolmogorov energy spectrum \citep{marshall2011development}.

\subsection{Optical Diagnostic}
\begin{figure}	
\centering
\includegraphics[width=0.7\textwidth]{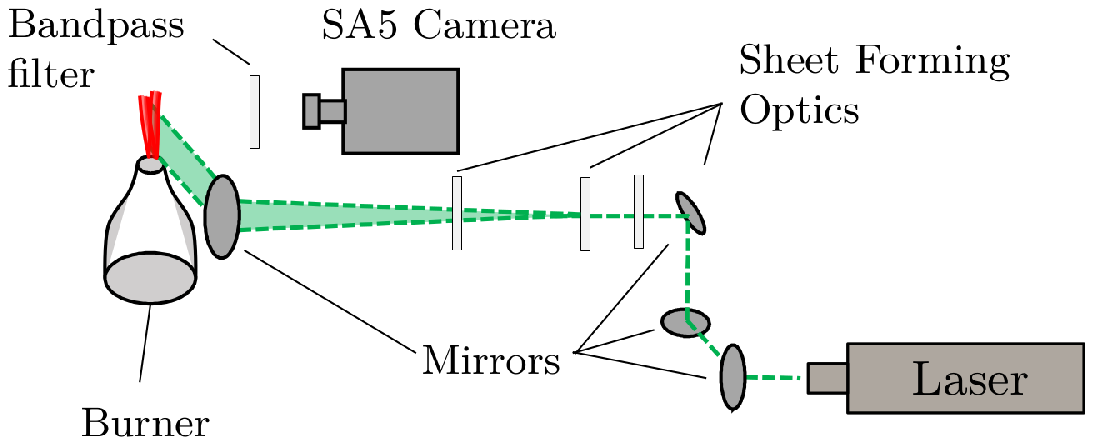}
\caption{\label{FigS2_OpticalDiagnostic} Arrangement of the optical diagnostic setup for measuring the flame surface and velocity field. Adapted from \cite{humphrey2018premixed} with permission from Cambridge University Press.}
\end{figure}

The optical diagnostic setup used for simultaneously measuring the flame dynamics and the turbulent flow is shown in Fig. \ref{FigS2_OpticalDiagnostic}. The flame edge is detected using $\text{TiO}_2$ Mie scattering, and the velocity field is quantified through particle image velocimetry (PIV). Flame images are acquired using Photron Fastcam SAS high-speed video camera with a Nikon Micro-Nikkor \textit{f} $= 55$m \textit{f}/$2.8$ lens. For the experiments, the resolution was set at $640\times 848$ pixels. The camera and laser pulse are controlled together by a dual head and are triggered together by a timing box. The laser used for diagnostics is a frequency-doubled Litron Nd:YLF with $527$ nm wavelength. The sampling frequency of Mie scattering imaging was kept fixed at $f_s = 1.25\times 10^4$ Hz to eliminate spectral leakage and bias errors in spectral estimation. In total, 21094 images were obtained for 1.68 s for each of the two flame configurations considered here.

The flow is seeded with Titanium dioxide ($\text{TiO}_2$) particles having a nominal diameter of 1 $\mu$m. The seeding is achieved by a cyclone seeder through which a portion of the main air is diverted before the premixing plenum. The seeded flow enters upstream of the settling chamber, as can be seen in Fig. \ref{FigS1_ExptSetup}. Only the main flow is seeded.

LaVision DaV PIV software \citep{LaVision} is used to process the PIV using a multipass algorithm. The first pass of which uses a 48$\times$48 pixel interrogation window with a 25\% overlap between windows. The subsequent passes use an 8$\times$8 pixel window, with 25\% overlap. This results in a resolution of 6 pixels ($\sim$ 0.46 mm) between vectors. For a description of uncertainty in PIV measurements, kindly refer to \citep{humphrey2018premixed, humphrey2017ensemble}. 

\begin{figure*}
\centering
\includegraphics[width=0.8\textwidth]{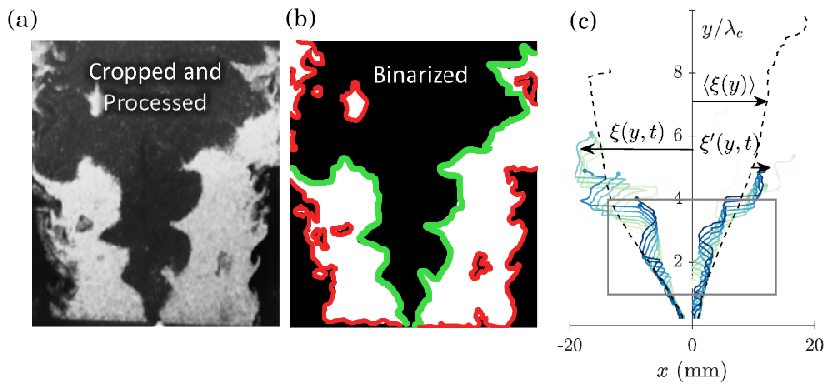}
\caption{\label{FigS3_ImageProcessing} (a) Cropped and processed Mie scattering image of the flame. (b) Binarized flame image following Otsu's method \citep{otsu1979threshold}. (c) Single-valued instantaneous $\xi(y,t)$, mean $\langle\xi(y)\rangle$ and fluctuating $\xi^\prime(y,t)$ flame surface. The flame edges here represent the leading edge, which is described in the text. Panel a,b have been adapted from \citep{humphrey2018premixed} with permission from Cambridge University Press.}
\end{figure*}

\begin{figure}
\centering
\includegraphics[width=0.45\textwidth]{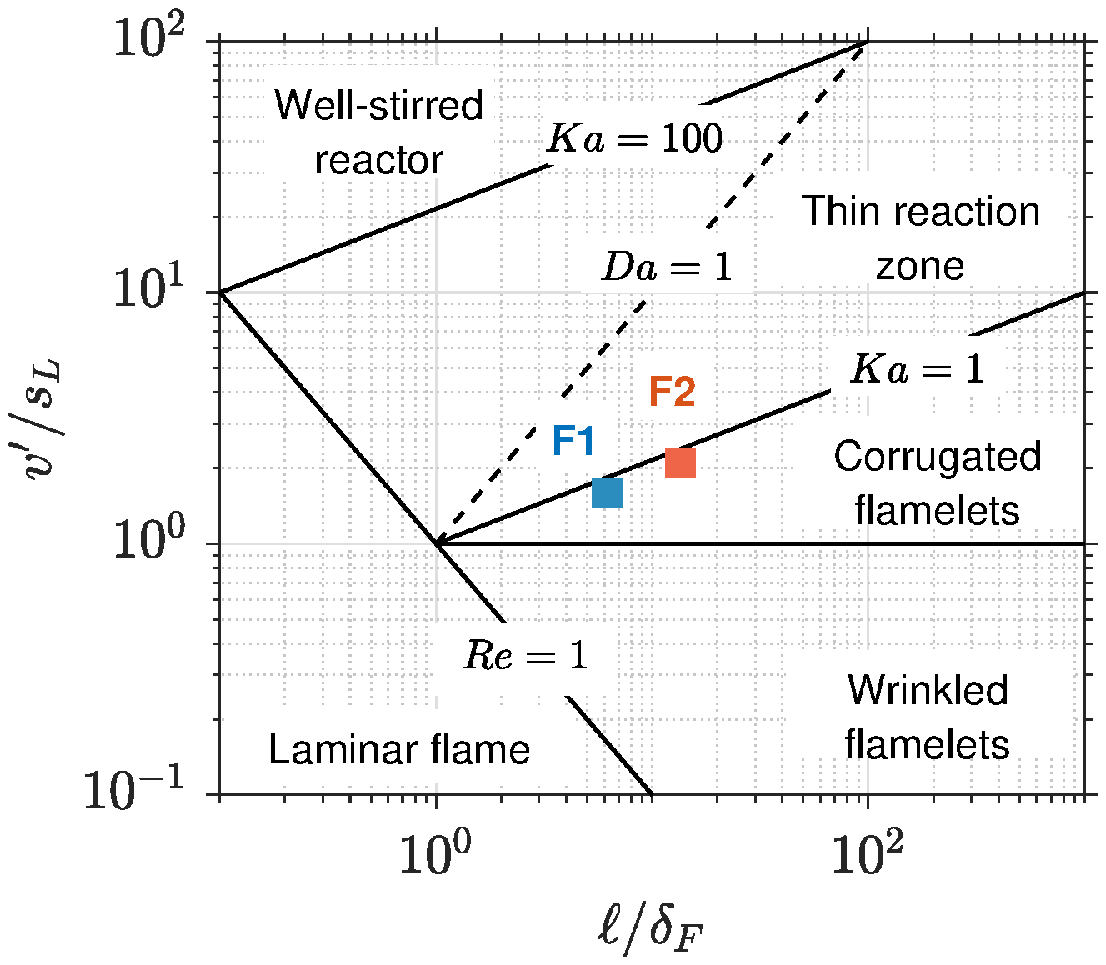}
\caption{Turbulent premixed flame regime diagram indicating the properties of Flame F1 and F2 considered in the present study.}
\label{Fig-Regime_diagram}
\end{figure}

\section{Flame edge detection}
\label{sec-S2}
Raw images acquired during the experiments are de-wrapped using LaVision DaVis PIV processing software to remove flame distortions due to the presence of the glass window. Figure \ref{FigS3_ImageProcessing}(a) shows a representative flame image that has been cropped and processed. The Mie scattering images are then binarized using a weighted threshold based on Otsu's method \citep{otsu1979threshold}. Figure \ref{FigS3_ImageProcessing}(b) shows the resultant binarized image.  The identified flame front from the binarized edge is indicated in green in Fig. \ref{FigS3_ImageProcessing}(b). As the flame-flow interaction lies in the limit of corrugated flamelets and thin reaction zone (cf. Fig. \ref{Fig-Regime_diagram}), the flame front remains continuous \citep{chowdhury2017experimentala}, allowing for a well-defined description of the flame front. We extract the instantaneous flame edge $\widetilde{\xi}(x,y,t)$, while ignoring any flame holes or islands that may be present. The mean and instantaneous flame edges are shown in Fig. \ref{FigS3_ImageProcessing}(c), where $x$ indicates the spanwise and $y$ indicates the streamwise direction. 

\begin{figure}
\centering
\includegraphics[width=0.75\textwidth]{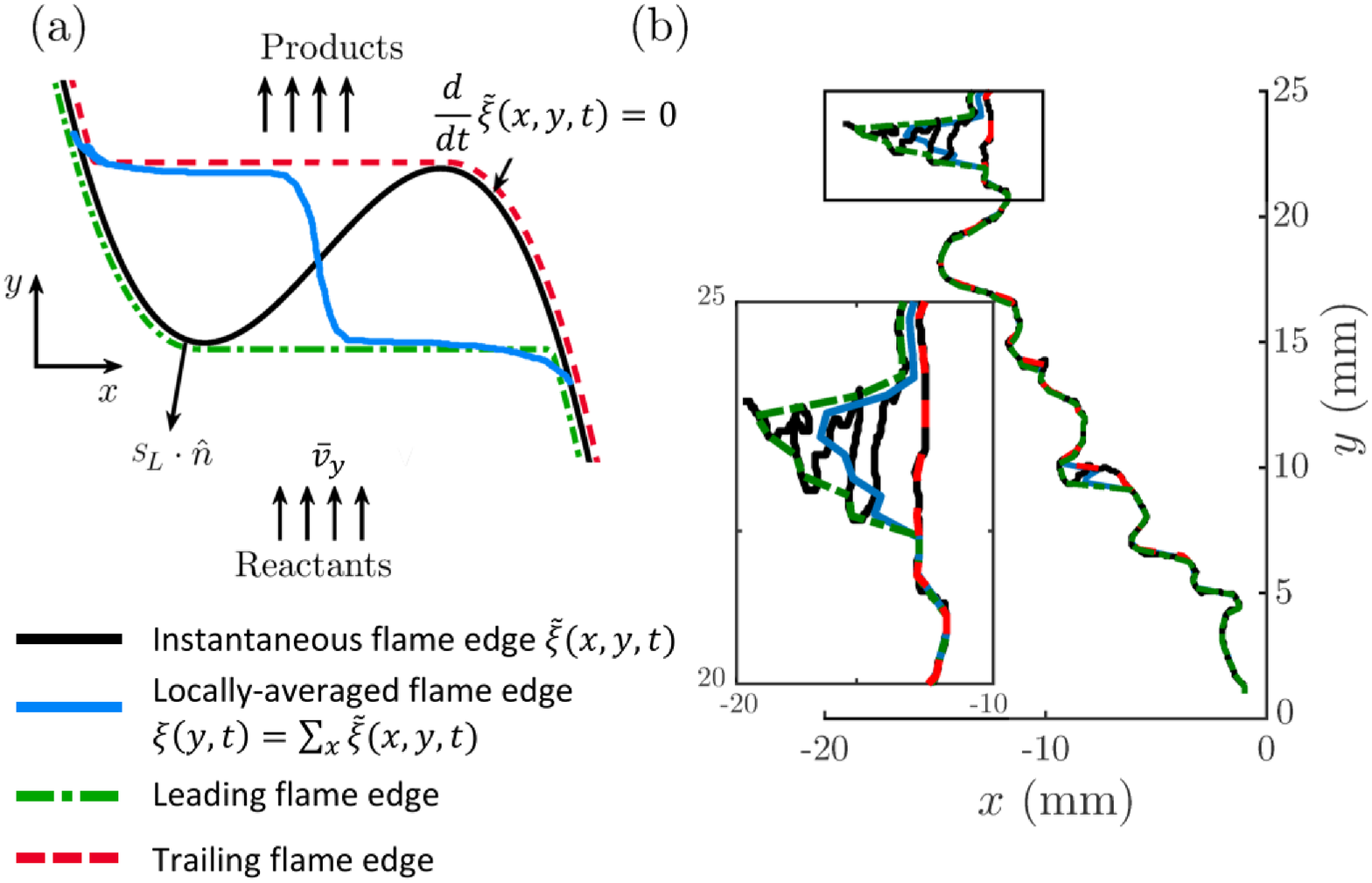}
\caption{\label{FigS4_FlameLeadingTrailingResponse}(a) Exaggerated pictorial depiction of converting multi-valued flame front $\widetilde{\xi}(x,y,t)$ into a locally averaged, leading and trailing flame edge $\xi(y,t)$. These single-valued flame edges are shown for a representative flame edge in (b).}
\end{figure}

Here, $\widetilde{\xi}(x,y,t)$ can be multivalued, i.e.,  for a given axial location there are multiple flame positions. We extract a single-valued flame position, $\xi(y,t)$ by three different strategies to obtain (i) a locally-averaged flame edge, (ii) a leading flame edge and (iii) a trailing flame edge, as follows:

\begin{enumerate}
\item We first calculate the time-averaged flame position:
\begin{equation}
\langle\xi(y)\rangle=\frac{1}{N_t}\sum_t \frac{1}{N_x(t)}\sum_{x(t)} \widetilde{\xi}(x,y,t),
\label{Eq1-MeanFlameEdge}
\end{equation}
where $N_t$ refers to the number of images in the time series, and $N_x$ refers to the number of multi-valued flame locations in the $x$ direction for a given $y$ location at a given instant of time. Thus, if the flame is not multi-valued at a given time instant, $N_x=1$.

\item We define the instantaneous, \textit{locally averaged} flame edge, by averaging over all the $x$-locations at which the flame is multi-valued, as
\begin{equation}
\xi(y,t) = \frac{1}{N_x} \sum_x \{\widetilde{\xi} (x,y,t)\}.
\label{Eq2-SpatiallyAveraged}
\end{equation}

\item We define the \textit{leading} and \textit{trailing} flame edges as
\begin{equation}
\xi(y,t) = \left\{\begin{array}{lr}
\displaystyle\sup\{\widetilde{\xi} (x,y,t)\}, \enspace \text{Leading Edge}\\
\\
\displaystyle\inf\{\widetilde{\xi} (x,y,t)\}, \enspace   \text{Trailing Edge}
\end{array}\right.
\label{Eq3-LeadingTrailingEdge}
\end{equation}
here, $\sup$ and $\inf$ are the \textit{supremum} and \textit{infimum} of the set $\{\widetilde{\xi}(x,y,t)\}$. Thus, the leading and trailing flame edge are the farthest and closest points on the flame front from the $y$-axis at every streamwise location, respectively, while the locally averaged flame edge resides at an intermediate position between these extremes.

\item Finally, the fluctuations are determined as $\xi^\prime(y,t) = \xi(y,t) - \langle\xi(y)\rangle$.
\end{enumerate}
For the single-valued flame front, all three flame edges are identical. In the case of multi-valued edges, the leading edge propagates into the reactants before the locally averaged flame front. In contrast, the trailing edge propagates into the reactants after the locally averaged flame front (see inset of Fig. \ref{FigS4_FlameLeadingTrailingResponse}b). The $x$-averaged edge lies in between the other two and has the effect of smoothing out artificial abrupt variations that may arise in the other two flame edges due to multi-valued folds. Our analysis of inner intermittency in the main text uses the leading flame edge definition specified in \eqref{Eq3-LeadingTrailingEdge}. In addition, we have also used the definition of the locally averaged flame edge defined in \eqref{Eq2-SpatiallyAveraged} in Appendix A to show that the anomalous scaling exponent remains unchanged when definition \eqref{Eq2-SpatiallyAveraged} is used for the intermittency analysis instead of \eqref{Eq3-LeadingTrailingEdge}. Appendix A also includes an additional test which further confirms that anomalous scaling and the inner intermittency it evidences are not an artifact of the representation of multi-valued wrinkles but a genuine feature of the flame dynamics.

\section{Properties of the turbulent flow field and the flame}

The instantaneous flow field $\bm{u}(\bm{x},t)$ is obtained from the PIV measurements. The mean $\langle \bm{u}\rangle$ and fluctuating $\bm{u}^\prime$ components of the velocity field are defined as
\begin{equation}
\bm{u}(\bm{x},t)=\langle \bm{u}(\bm{x})\rangle +\bm{u}^\prime(\bm{x},t),
\end{equation}
where, $\langle\cdot\rangle$ denotes an average over time, so that $\langle\bm{u}(\bm{x})\rangle =1/T\int_T \bm{u}(\bm{x},t)dt$. The averaging time $T(\geq 250\tau_\ell)$ is several times greater than the integral time $\tau_\ell$ of the flow. The turbulent intensity is measured as: $\upsilon=\langle 1/3(\bm{u}^\prime \cdot\bm{u}^\prime)\rangle$. Further, we define the velocity auto-correlation function as:
\begin{equation}
R(\bm{r})=\langle\bm{u}^\prime(\bm{x+r})\bm{u}^\prime(\bm{x})\rangle.
\end{equation}
The longitudinal $f(r)$ and lateral $g(r)$ velocity correlation functions, normalized by the kinetic energy,  are defined as
\begin{align}
f(r) = \left\langle u_i(\bm{x}+r\hat{\bm{e}}_i)u_i(\bm{x}) \right\rangle/u^2,\\
g(r) = \left\langle u_j(\bm{x}+r\hat{\bm{e}}_i)u_j(\bm{x}) \right\rangle/u^2.
\end{align}
where, $\hat{e}_i$ indicates a unit normal along the index $i$. 

\begin{figure*}
\centering
\includegraphics[width=0.8\textwidth]{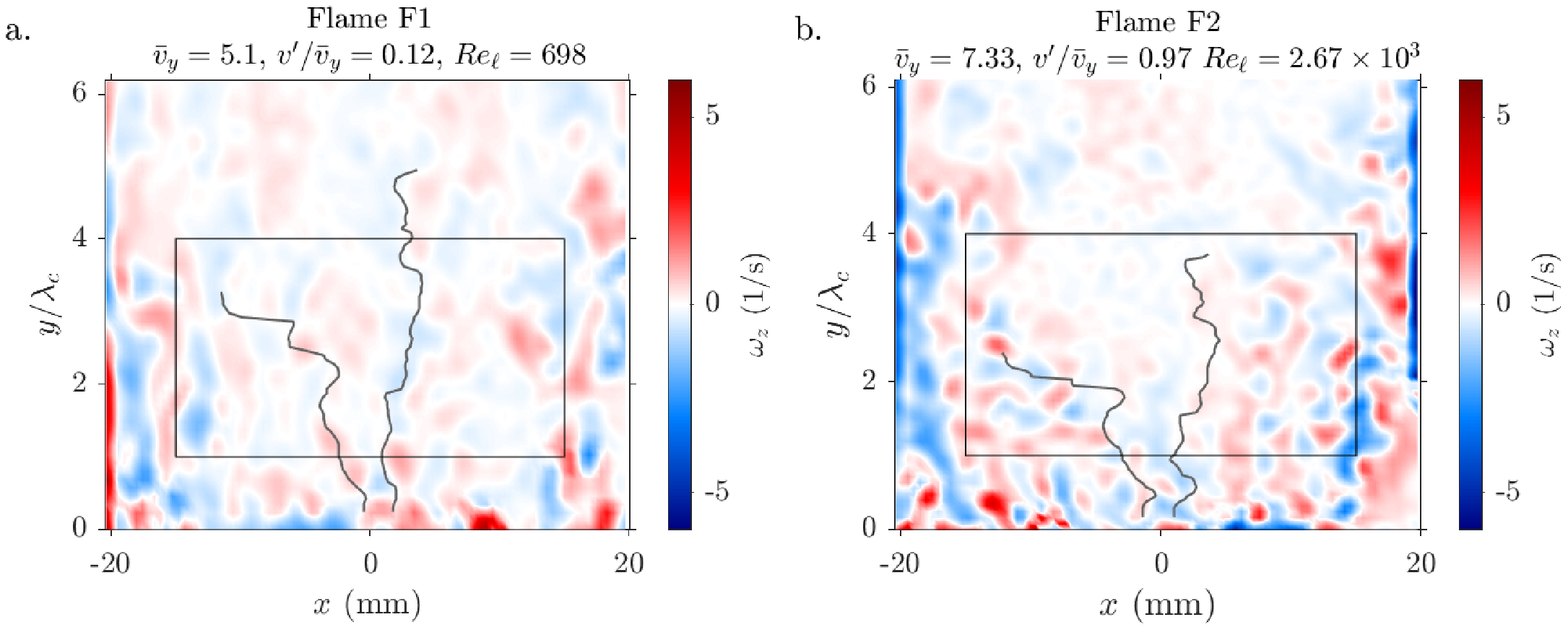}
\caption{Depiction of the instantaneous flow-field for the two flame configurations considered in this study. The boxed region indicates the local region chosen for measuring velocity statistics for the experimental cases discussed here.}
\label{FigS5}
\end{figure*}

\begin{figure}
\centering
\includegraphics[width=0.8\textwidth]{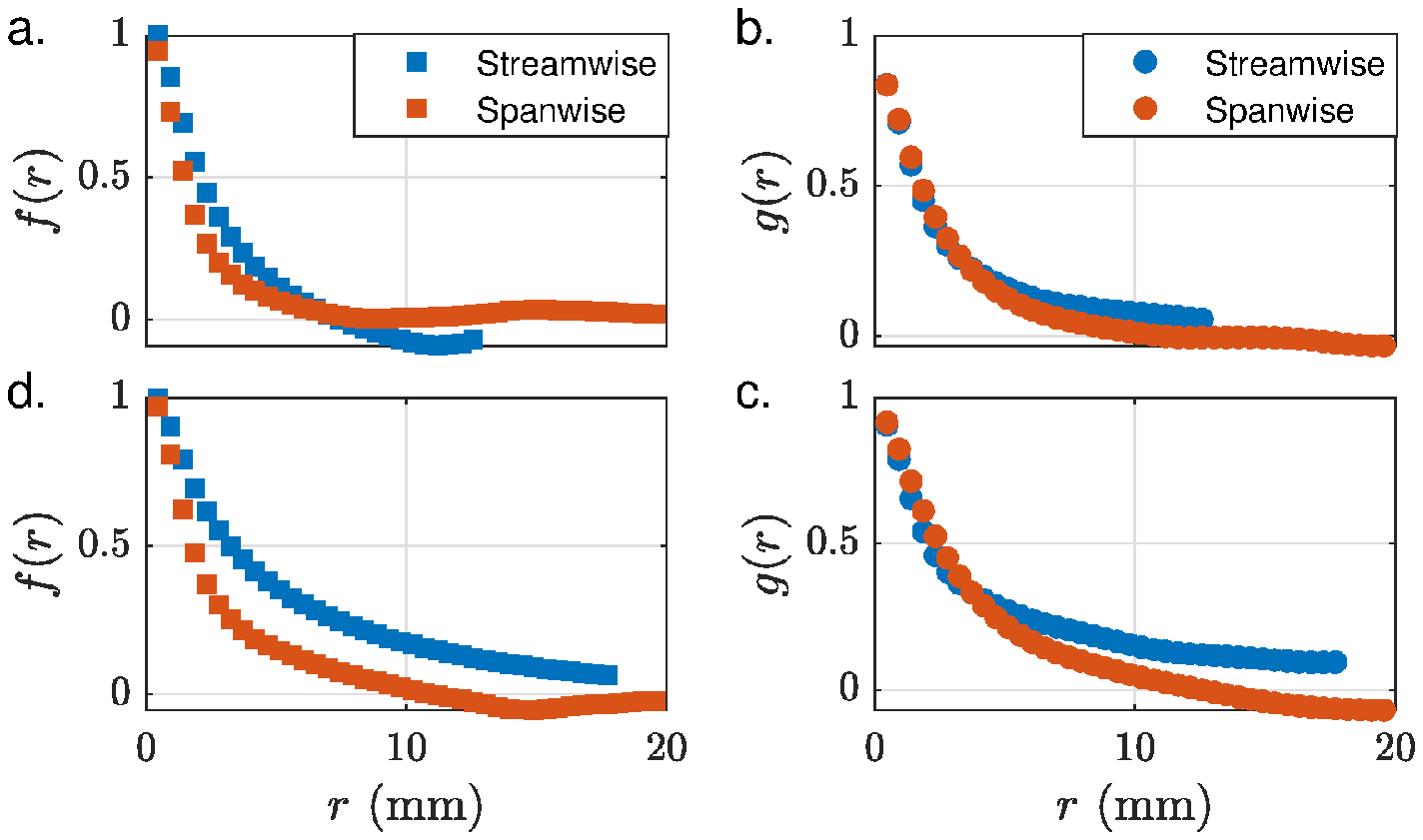}
\caption{Longitudinal ($f$) and lateral ($g$) velocity correlation as a function of $r$ (in mm) for flame configuration F1 (a,b) and F2 (c,d). The blue marker indicates measurement along the streamwise $y$-direction while the red marker indicates the measurement in the spanwise $x$-directions.}
\label{Fig-Velocity_correlation}
\end{figure}

\begin{figure*}
\centering
\includegraphics[width=0.8\textwidth]{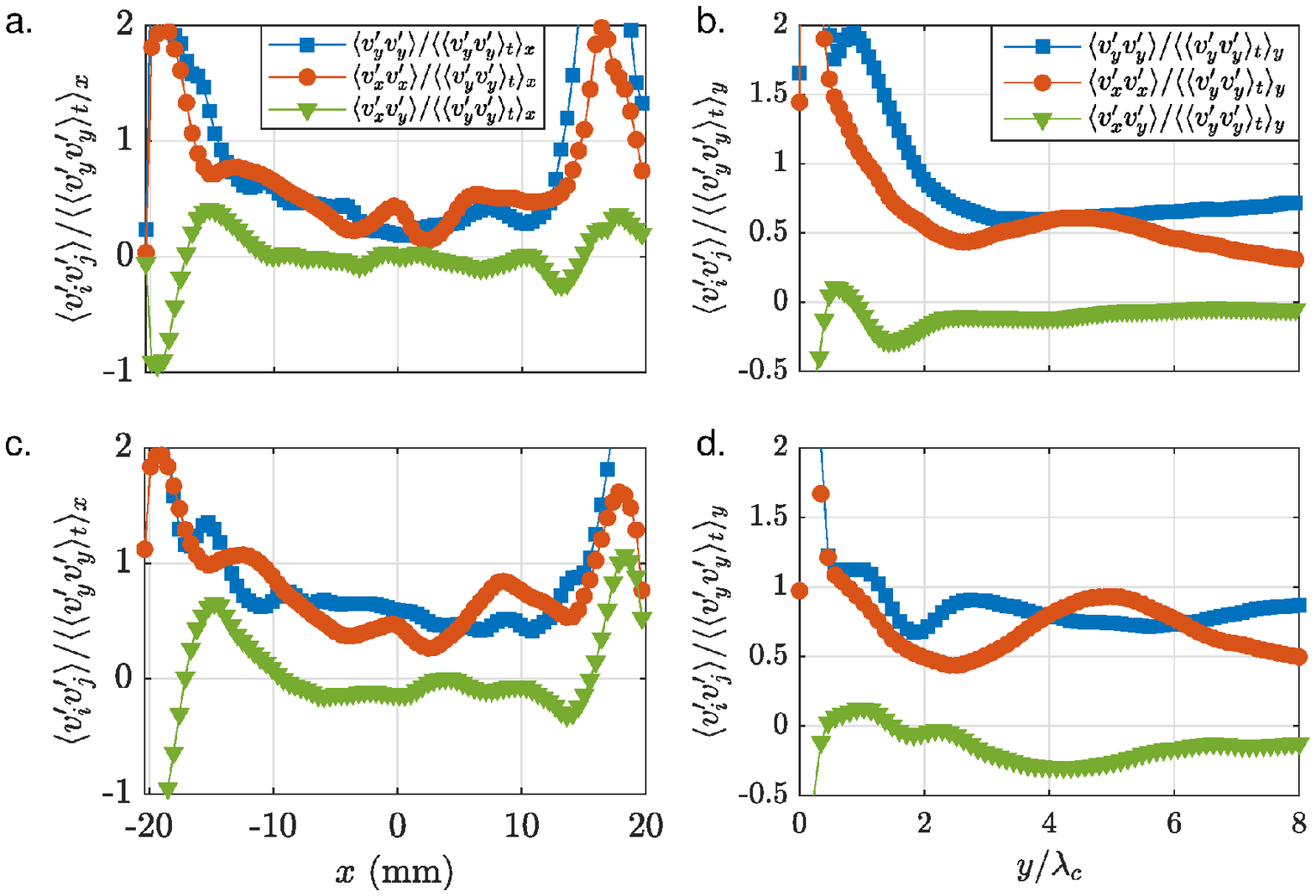}
\caption{Time-averaged distribution of velocity cross-correlation as a function of (a,c) spanwise direction $x$ measured at $y=2\lambda_c$ and (b, d) streamwise direction $y/\lambda_c$ at $x=-5$ mm for flame F1 (a, b) and F2 (c, d). The cross-correlation $\langle u^\prime_xu^\prime_y\rangle$ almost vanishes inside the boxed region indicated in Fig. \ref{FigS5}.}
\label{Fig-IsotropyLevel}
\end{figure*}

The longitudinal and lateral correlation functions for the two flame configurations F1 and F2, measured using data from the boxed region in Fig.~\ref{FigS5}, are presented in Fig. \ref{Fig-Velocity_correlation}. Here, the blue (red) markers correspond to the correlation function measured along the streamwise (spanwise) direction. We note that for both F1 (panel a, b) and F2 (panels c,d) the correlations measured in the two directions are quite similar. Moreover, the velocity cross-correlations $\langle u^\prime_x u^\prime_y \rangle$, shown in Fig.~\ref{Fig-IsotropyLevel} for the two flames, have relatively small values  ($-0.2<\langle u^\prime_xu^\prime_y\rangle<0.1$) for $-15<x<15$ mm and $\lambda_c<y<4\lambda_c$ (the boxed region in Fig.~\ref{FigS5}). These measurements indicate that the turbulent flow is close to isotropic within the local region in which we analyze the flame fluctuations.

The longitudinal correlations functions allow us to determine the integral and dissipative length scales of the flow. First, the integral length scale $\ell$ is determined as
\begin{equation}
\ell = \int_0^\infty f(r) dr.
\label{EqA5 - Integral_Scale}
\end{equation}
The integral scale Reynolds number is then determined as
\begin{equation}
Re_\ell = \frac{\rho u^\prime\ell}{\mu},
\label{EqA6-Turbulent_Reynolds_Number}
\end{equation}
where, the properties of the binary $CH_4/$air mixture $\rho$ and $\mu$ are obtained following \citep{wilke1950viscosity}.

Next, the rate of turbulent kinetic energy dissipation is obtained from the large-scales as: $\varepsilon\sim \upsilon^3/\ell$. This finally allows us to determine the dissipative, Kolmogorov length $\eta$ and time $\tau_\eta$ scales in terms of the integral scale quantities:
\begin{equation}
\eta/\ell=Re_\ell^{-3/4}, \qquad \tau \upsilon/\ell=Re_\ell^{-1/2}.
\label{EqA7-Kolmogorov}
\end{equation}

To determine the key length and time scales of the flame, we begin by calculating the laminar flame speed $s_L$ via Chemkin PREMIX calculations \citep{kee2011chemkin}, using detailed chemistry simulated through GRIMech 3.0 mechanism \citep{smith1999gri} at 300 K and 1 bar. The flame thickness is calculated using the temperature gradient between the unburned reactants and burnt products, to wit, 
\begin{equation}
\delta_F=\frac{T^u-T^b}{\max(dT/dx)}.
\end{equation}

The non-dimensional Damkohler (Da) and Karlovitz (Ka) numbers are then determined as
\begin{equation}
Da = \tau/\tau_{\text{chem}}=\left(\eta/\delta_F\right)^2, \qquad Ka = 1/Da.
\label{EqA9-Damkohler-Karlovitz}
\end{equation}
The Damk\"ohler number of flame F1 is $Da=3.91$ and F2 is $Da=5.39$, and correspond to the boundary of corrugated flamelets and thin reaction zone (see Fig. \ref{Fig-Regime_diagram}).

The important flame length scales are the Gibson length scale ($\ell_g$) and the Corrsin length scale ($\eta_c$). These are defined as
\begin{equation}
\ell_g = \left(s_L/\upsilon\right)^3\ell, \qquad \eta_c = (\mathcal{D}_L^3/\varepsilon)^{1/4}=Sc^{-3/4}\eta.
\label{EqA10-Gibson-Corrsin-Length-Scale}
\end{equation}
Here, $\mathcal{D}_L$ is the Markstein diffusion. In determining the Corrsin scale $\eta_c$, we have taken the Schmidt number $Sc=\nu/\mathcal{D}_L=0.7$ for both flame configurations following \cite{tamadonfar2014flame}. The properties of the two turbulent flames configurations have been tabulated in Table \ref{tab:Expt Features}.

\begin{table}
  \begin{center}
\def~{\hphantom{0}}
\begin{tabular}{lcc}
\textbf{Feature} & \textbf{Flame F1} & \textbf{Flame F2} \\[3pt]
      $\bar{u}_y$ & $5.1$ m/s & $7.33$ m/s\\
      $u^\prime$ & $0.59$ m/s  & $0.97$ m/s\\
      $u^\prime/\bar{u}_y$ & $0.12$ & $0.13$ \\
      $\phi$ & $0.97$ & $0.91$ \\
      $\nu$ & $2.29\times10^{-6}$ m$^2$/s & $2.27\times10^{-6}$ m$^2$/s\\
      $f_f$ & $1250$ Hz & $1250$ Hz\\
      $\lambda_c=\bar{u}_y/f_f$ & $4.08$ mm & $5.9$ mm \\
      $Re_d=\bar{u}_yd/\nu$ & $7.02\times 10^3$ & $1.17\times 10^4$ \\
      $\ell$ & $2.72$ mm & $6.22$ mm\\
      $Re_\ell=u^\prime\ell/\nu$ & $6.98\times 10^2$ & $2.67\times 10^3$\\
      $s_L$  & $0.37$ m/s & $0.34$ m/s \\
      $\delta_F$ & $0.44$ mm & $0.46$ mm \\
      $Da=\ell s_L/\delta_Fu^\prime$ & $3.91$ & $5.39$\\
      $\eta=Re_\ell^{-3/4}\ell$ & $0.020$ mm & $0.017$ mm \\
      $\ell_g=(s_L/u^\prime)^3\ell$ & $0.68$ mm & $0.26$ mm\\
\end{tabular}
\caption{\label{tab:Expt Features} Relevant properties of the two turbulent premixed flames considered in this study. The quantities $s_L$ and $\delta_F$ were obtained using Chemkin Premix calculations \citep{humphrey2017ensemble}. See supplementary \S3 for details of the turbulent field.}
\end{center}
\end{table}

\section{Outer intermittency in large-scale flame fluctuations}
\begin{figure}
\centering
\includegraphics[width=0.7\textwidth]{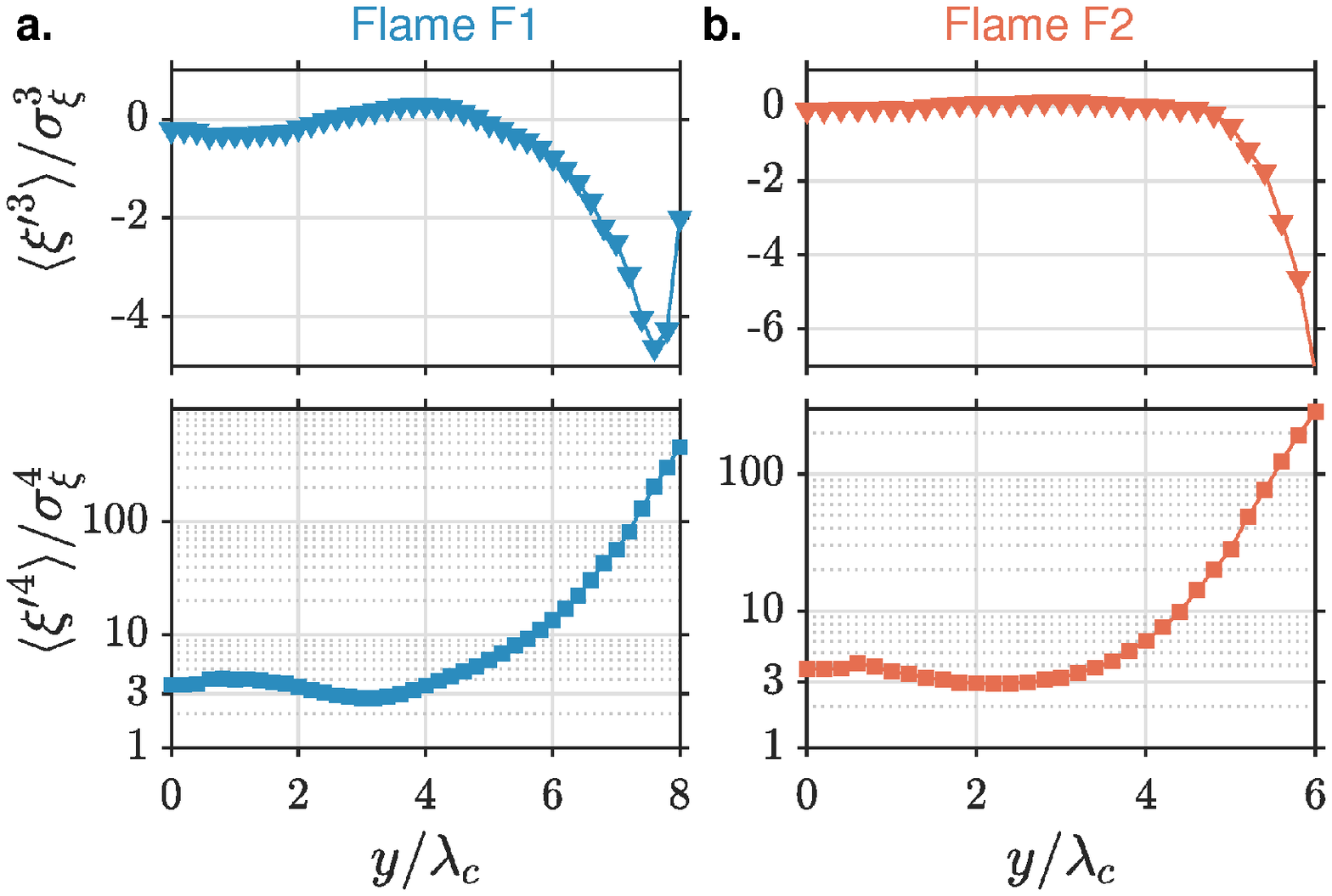}
\caption{Variation of the skewness $\langle{\xi^\prime}^3\rangle/\sigma_\xi^3$ (top) and kurtosis $\langle{\xi^\prime}^4\rangle/\sigma_\xi^4$ (bottom) of flame fluctuations $\xi^\prime$ as a function of downstream location from flame holder $y/\lambda_c$ for Flames F1 and F2.}
\label{FigS8}
\end{figure}

In Fig. 1(c,d) in the manuscript, we observed that the flame fluctuations, at large distances from the flame holder, depict an ``on-off" type intermittent flapping behavior. To characterize this further, we plot the skewness and kurtosis of flame fluctuations as a function of $y/\lambda_c$ for the two flames in Fig. \ref{FigS8}. We can observe that while the kurtosis $K\approx 3$ and the skewness $Sk\approx 0$ for small distances 
($0\leq y\leq 4$), both properties depart dramatically from these Gaussian values as we move away from the flame holder. The large kurtosis implies the outer-intermittency of the large-scale flame motion, discussed in the main text. Note also that both flames depict negative skewness, implying that the large-scale flapping fluctuations are more likely to be in the direction of the unreacted fuel-air mixture than in the direction of the products.

\begin{figure}
\centering
\includegraphics[width=0.5\textwidth]{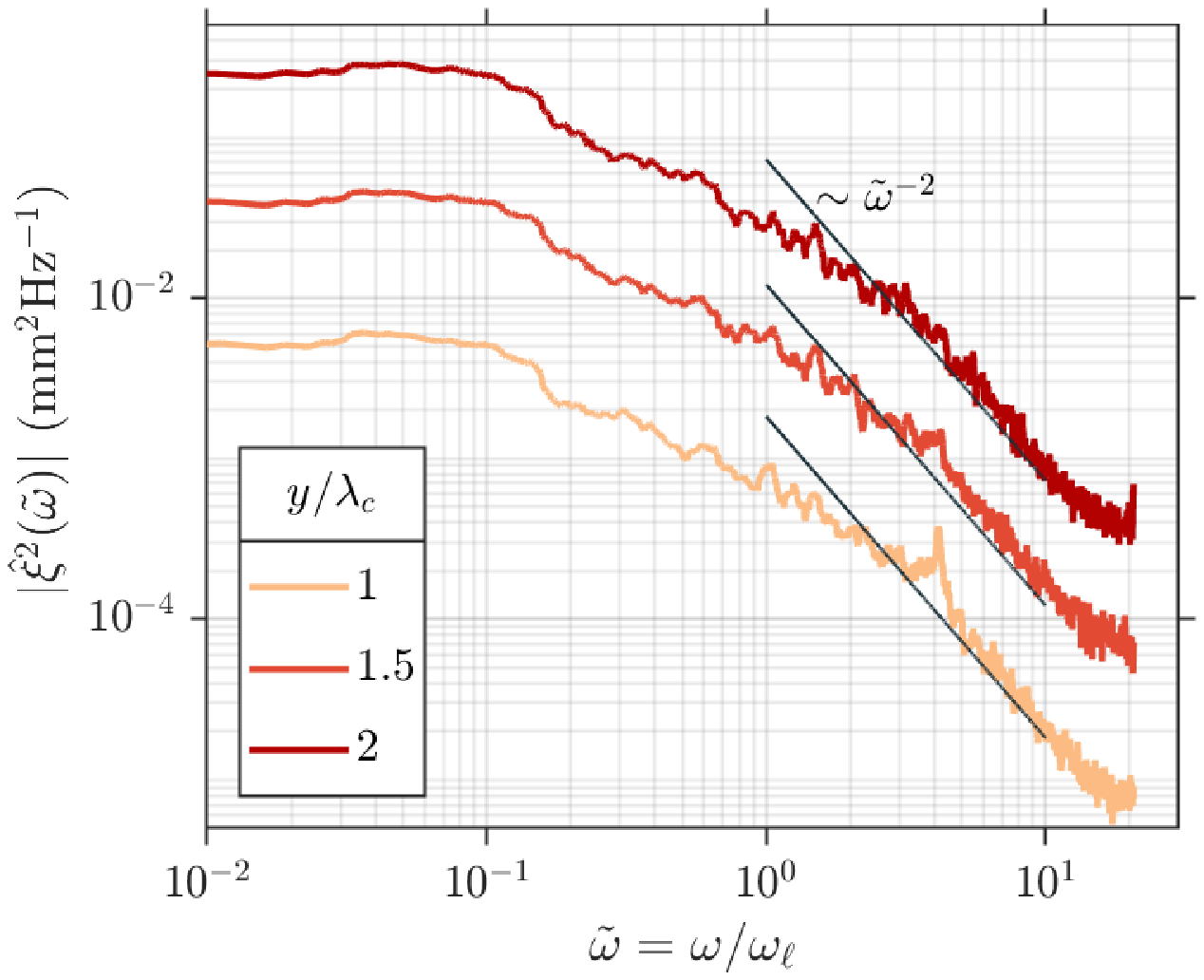}
\caption{Temporal spectrum of flame fluctuations for Flame F2 measured at the indicated $y/\lambda_c$ locations. For measurements at locations $y/\lambda_c<1$, the flame response is harmonic with a peak at the forcing frequency $\tilde{\omega}_f$.}
\label{FigS9}
\end{figure}

\begin{figure}
\centering
\includegraphics[width=0.6\textwidth]{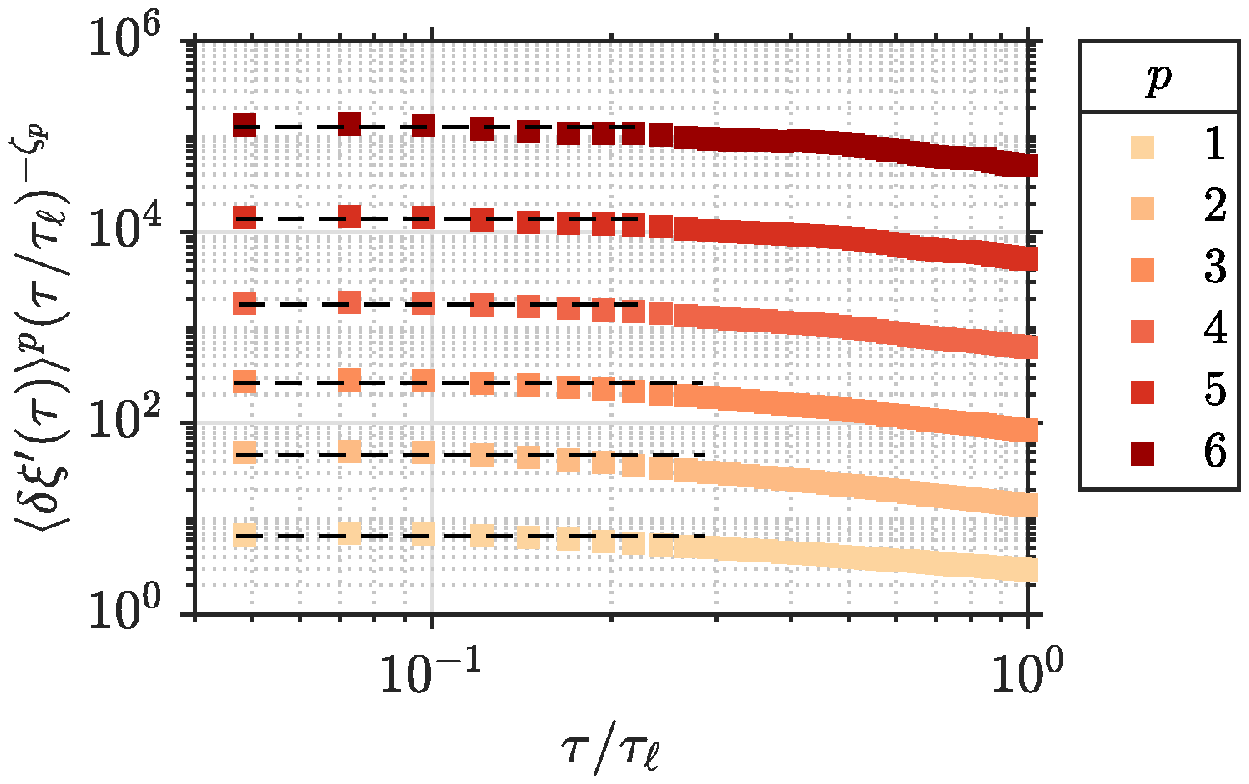}
\caption{Structure function compensated by the estimated scaling $\tau^{-\zeta_p}$ for various order $p$ as a function of the time delay $\tau$ for Flame F2.}
\label{FigS10}
\end{figure}

\section{Scaling of power-spectrum and higher-order structure-function of Flame F2}

The scaling of the power spectrum for flame F1 was depicted in Fig. 2 of the manuscript. The fluctuations associated with flame F2 also depict the same scaling behavior with a scaling exponent $\alpha \approx -2$ for measurements made between $1 \leq y/\lambda_c\leq3$. This can be seen in Fig. \ref{FigS9}. As was the case with F1, we observe that fluctuations measured at locations close to the flame holder depict a narrowband peak at the forcing frequency. However, for $y/\lambda_c>1$, the effect of narrowband forcing gives way to a power-law scaling with $\alpha=-2$. The variation of $\alpha$ with $y/\lambda_c$ is plotted, for both flames F1 and F2, in the inset in Fig. 2 of the main manuscript. 

The compensated structure-function of the flame fluctuations for flame F2 is shown in Fig. \ref{FigS10}. We notice that the structure functions depict well-defined scaling behavior where the scaling lasts for over one decade. The structure-functions are compensated by their scaling exponent $\zeta_p$ up to order $p=6$. The variation of $\zeta_p$ with $p$, for both flames F1 and F2, has been shown in Fig. 3(a) of the manuscript.




\begin{figure}
\centering
\includegraphics[width=0.9\textwidth]{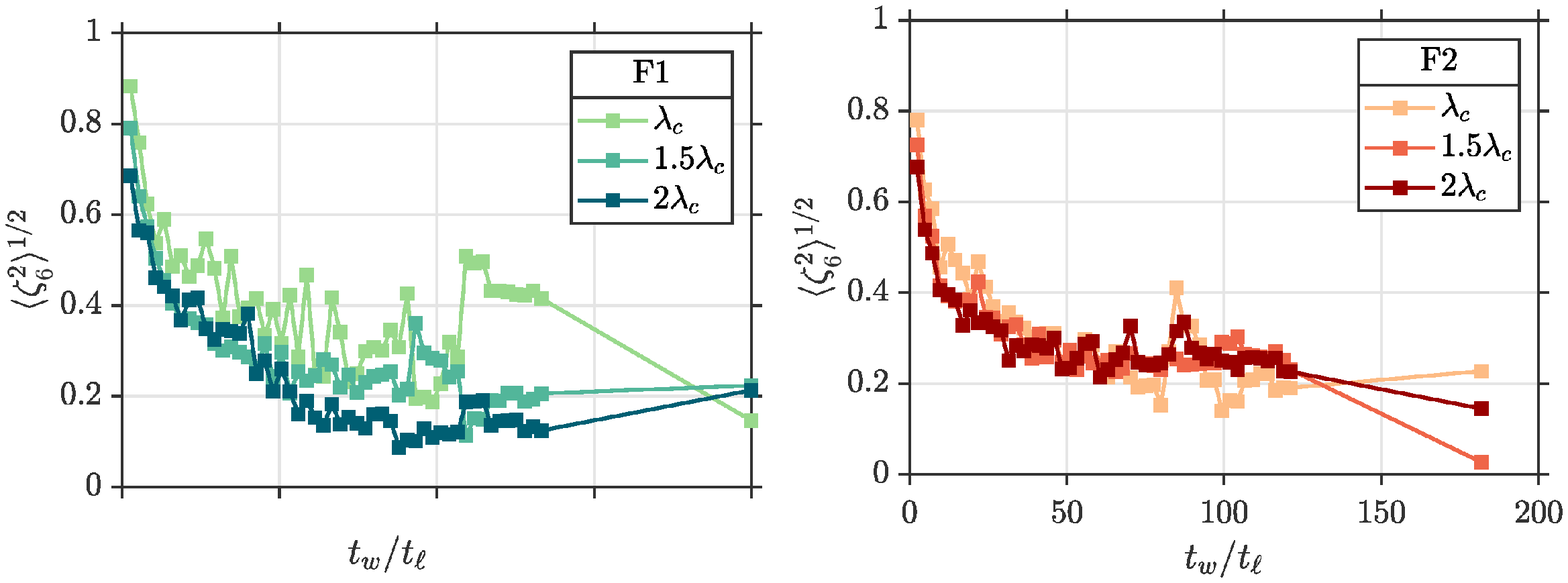}
\caption{\label{FigS12} The variation in the standard deviation of $\zeta_p$ for $p=6$ for various window sizes $t_w$ at various $y/\lambda_c$ locations for the two flames. We have chosen $t_w=100t_\ell$ for obtaining the errorbars in Fig. 3a of the manuscript.}
\end{figure}

\section{Statistical uncertainty in the measurement of scaling exponents}

In this section we describe how we estimated the error-bars associated with the scaling exponents $\zeta_p$, shown in Fig.~3a in the main manuscript. For each flame configuration, the available data set consists of 21094 snapshots of the flame (at a sampling frequency $f_s=1.25\times10^4$ Hz). Considering the azimuthal symmetry of the combustor setup, we calculated the flame fluctuations at a given $y$ location from both the left and right flame edges, to obtain 42188 data points. This data set was then divided into subsets, each of which spanned a time window of size $t_w$. Each subset of data was then used to calculate the flame position increments, the structure functions and the scaling exponents. Thus, we obtain several values of $\zeta_p$, one for each of the subsets of data. The mean of these values $\langle \zeta_p \rangle$ is reported as the measured values of the scaling exponent, while the standard deviation of these values $\langle \zeta_p^2 \rangle^{1/2}$ is used to estimate the associated statistical uncertainty.

To aid in selecting the size of the time window $t_w$, we calculate the standard deviation of the scaling exponents for various $t_w$. An illustrative example is depicted in Fig.~\ref{FigS12}, for the scaling exponent of the sixth-order structure-function. We see that for both flame F1 and F2, the standard deviation decreases as $t_w$ is increased up to about $t_w=80 t_\ell$, beyond which the standard deviation saturates. So we choose $t_w=100 t_\ell$ as the time window and report the corresponding mean and standard deviation as the markers and error bars in Fig.~3a of the main manuscript.

\bibliographystyle{jfm} 
\bibliography{references_sup}